\documentclass[journal]{IEEEtran}

\IEEEoverridecommandlockouts

\ifCLASSINFOpdf
\else
\fi

\usepackage{cite}
\usepackage{amsmath,amssymb,amsfonts}
\usepackage{algorithmic}
\usepackage[linesnumbered,algoruled,boxed,lined]{algorithm2e}
\usepackage{graphicx}
\usepackage{textcomp}
\usepackage{xcolor}
\usepackage{multirow}
\usepackage{bm}
\usepackage{booktabs}
\usepackage{ntheorem}
\usepackage{ulem}
\usepackage{tabularx}
\usepackage[caption=false]{subfig}

\theoremseparator{.}
\newskip\theorempreskipamount
\newskip\theorempostskipamount

\newtheorem{lemma}{Lemma}

\theorembodyfont{}

\newenvironment{proof}{{\noindent\it Proof. }}{\hfill $\blacksquare$\par}

\allowdisplaybreaks[4]

\begin{document}
\pagestyle{empty} 

\title{Collaborative Ground-Space Communications via \\ Evolutionary Multi-objective Deep \\ Reinforcement Learning}

\author{Jiahui Li, \IEEEmembership{Student Member,~IEEE,}
        Geng~Sun, \IEEEmembership{Member,~IEEE,}
        Qingqing~Wu, \IEEEmembership{Senior Member,~IEEE,} \\
        Dusit Niyato, \IEEEmembership{Fellow,~IEEE,} 
        Jiawen Kang,~\IEEEmembership{Member,~IEEE,} 
        Abbas Jamalipour,~\IEEEmembership{Fellow,~IEEE,} \\
        and Victor C. M. Leung,~\IEEEmembership{Life Fellow,~IEEE}
        % <-this % stops a space
	% \thanks{This study is supported in part by the National Natural Science Foundation of China (62002133, 61872158, 61806083), in part by the Science and Technology Development Plan Project of Jilin Province (20190701019GH, 20190701002GH), and in part by Youth Science and technology talent lift project of Jilin Province. }\protect

  \thanks{Jiahui Li is with the College of Computer Science and Technology, Jilin University, Changchun 130012, China (e-mail: lijiahui0803@foxmail.com). 
  \par Geng Sun is with the College of Computer Science and Technology, Jilin University, Changchun 130012, China, and also with the School of Computer Science and Engineering, Nanyang Technological University, Singapore 639798 (e-mail: sungeng@jlu.edu.cn). (\textit{Corresponding author: Geng Sun.)}
  \par Qingqing Wu is with the Department of Electronic Engineering, Shanghai Jiao Tong University, Shanghai 200240, China  (e-mail: qingqingwu@sjtu.edu.cn). 
  \par Dusit Niyato is with the School of Computer Science and Engineering, Nanyang Technological University, Singapore 639798 (e-mail: dniyato@ntu.edu.sg). 
  \par Jiawen Kang is with the School of Automation, Guangdong University of Technology, Guangzhou 510006, China (e-mail: kjwx886@163.com). 
  \par Abbas Jamalipour is with the School of Electrical and Computer Engineering, The University of Sydney, Sydney, NSW 2006, Australia (e-mail: a.jamalipour@ieee.org). 
  \par Victor C. M. Leung is with the College of Computer Science and Software Engineering, Shenzhen University, Shenzhen 518060, China, and also with the Department of Electrical and Computer Engineering, The University of British Columbia, Vancouver V6T 1Z4, Canada (e-mail: vleung@ieee.org). 
  }

}

\IEEEtitleabstractindextext{%
\begin{abstract}
In this paper, we propose a distributed collaborative beamforming (DCB)-based uplink communication paradigm for enabling ground-space direct communications. Specifically, DCB treats the terminals that are unable to establish efficient direct connections with the low Earth orbit (LEO) satellites as distributed antennas, forming a virtual antenna array to enhance the terminal-to-satellite uplink achievable rates and durations. However, such systems need multiple trade-off policies that variously balance the terminal-satellite uplink achievable rate, energy consumption of terminals, and satellite switching frequency to satisfy the scenario requirement changes. Thus, we perform a multi-objective optimization analysis and formulate a long-term optimization problem. To address availability in different terminal cluster scales, we reformulate this problem into an action space-reduced and universal multi-objective Markov decision process. Then, we propose an evolutionary multi-objective deep reinforcement learning algorithm to obtain the desirable policies, in which the low-value actions are masked to speed up the training process. As such, the applicability of a one-time trained model can cover more changing terminal-satellite uplink scenarios. Simulation results show that the proposed algorithm outmatches various baselines, and draw some useful insights. Specifically, it is found that DCB enables terminals that cannot reach the uplink achievable threshold to achieve efficient direct uplink transmission, which thus reveals that DCB is an effective solution for enabling direct ground-space communications. Moreover, it reveals that the proposed algorithm achieves multiple policies favoring different objectives and achieving near-optimal uplink achievable rates with low switching frequency. 
\end{abstract}

}

\maketitle

\IEEEdisplaynontitleabstractindextext
\IEEEpeerreviewmaketitle

% Section
% Introduction
%
\section{Introduction}

\par While terrestrial networks, including the fifth-generation (5G) networks and Wi-Fi, have undergone extensive research and deployment, the current network architecture still faces challenges in providing coverage in remote areas and exhibits fragility during natural disasters~\cite{Heo2023}. In this case, non-terrestrial networks based on airborne movable elements, such as unmanned aerial vehicles (UAVs) and airships, have gained significant attention~\cite{Mahboob2024}. Recently, with the advancements in manufacturing processes, satellites become integral components of network architectures instead of only traditional roles in positioning and remote sensing, thereby significantly enhancing Internet coverage and disaster response capabilities~\cite{Zhou2023,Luglio2022}. For instance, SpaceX develops the Starlink project to deliver global high-speed, low-latency broadband internet services~\cite{Ma2023}. Moreover, the third generation partnership project (3GPP) discussed the integration of satellite networks in Rel-18, including radio access networks, services, system aspects, core, and terminals~\cite{3GPP38821}. 
% Furthermore, the implementation of mobile edge computing in satellite networks also holds promise for improving the quality of service and facilitating collaborative computing offloading~\cite{Luglio2022}. 

\par Among various platforms, low Earth orbit (LEO) constellations, consisting of thousands of satellites, play a crucial role in satellite networks by offering advantages such as lower transmission delay compared to medium Earth orbit and geostationary Earth orbit satellites~\cite{Cao2023}. Leveraging LEOs has empowered various terrestrial devices to establish direct connections with satellite networks, which grants them extensive Internet access capabilities in remote areas~\cite{Wang2023}. However, some previously deployed terrestrial terminals may be energy-sensitive and equipped with coarse antennas. In such cases, the uplink transmission from these terminals to LEO satellites can be low-efficiency and only stable when the link distances are short. As such, the terminals have to switch satellites to connect frequently, resulting in the vexing problem of ping-pong handovers~\cite{Yang2023}. Thus, it is of significant importance to improve the terrestrial-satellite uplink quality for enabling ground-space direct communications.
 
\begin{figure}
    \centering
    \includegraphics[width=1\linewidth]{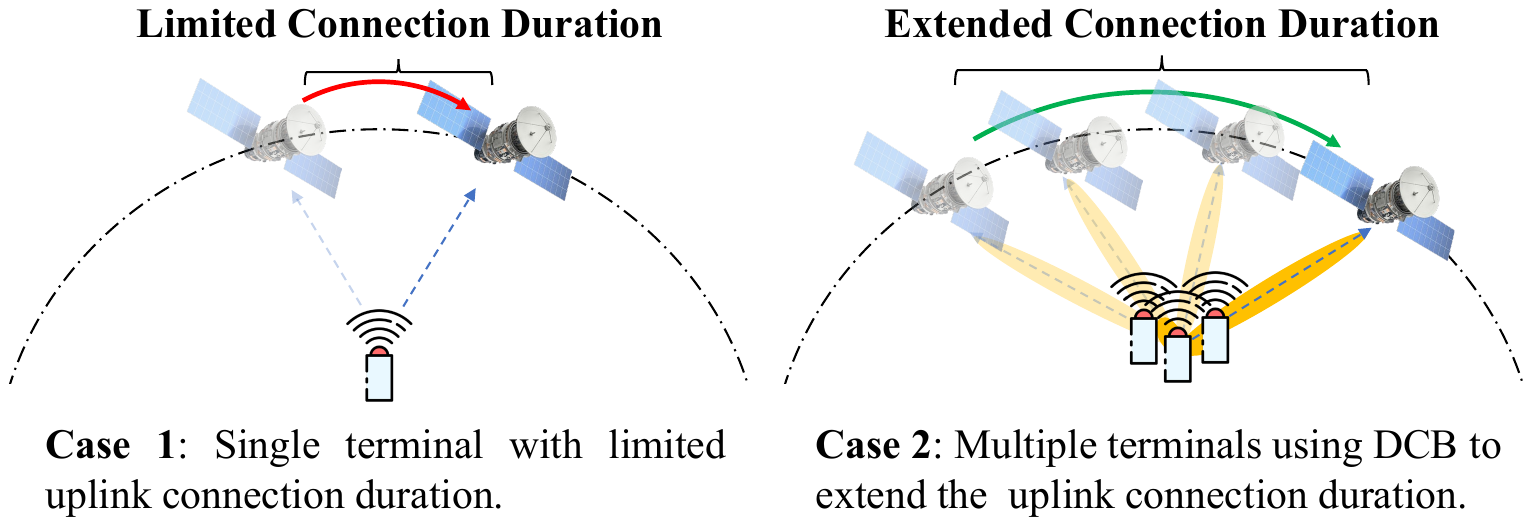}
    \caption{Due to the low uplink gain and transmit power of the terminals, the single terminal to LEO uplink only continues short time. Benefiting from the transmission gain of DCB, the virtual antenna array will achieve extended connection duration.} 
    \label{fig:duration}
\end{figure}

\par Distributed collaborative beamforming (DCB) can be introduced into terrestrial terminals to achieve this goal. Specifically, DCB treats separate systems such as these terminals as distributed antennas and simulates the beamforming process to produce a considerable transmission gain. This gain is beneficial to offset wireless fading even in long-range links between ground devices and satellites~\cite{Xu2023}, thereby enhancing the corresponding transmission distance and the received signal strength. In this way, as shown in Fig.~\ref{fig:duration}, we can adopt DCB to extend the connection duration of one satellite and enhance the uplink capabilities, thereby improving the uplink achievable rate and reducing the satellite switching frequency.

\par However, designing such a DCB-based terminal-to-satellite uplink communication system is a nontrivial task. \textit{First}, the uplink transmission performance and energy efficiency of DCB are determined by transmit power allocation of terminals. As such, transmit powers of terminals should be carefully optimized according to the channel conditions~\cite{Jung2021}. \textit{Second}, while DCB enhances transmission performance, the switching decision also needs to consider maximizing the uplink achievable rate and minimizing the satellite switching frequency. The relative importance of these two goals may vary across diverse scenarios, which means that the existing single-objective optimization and static methods in the literature (\textit{e.g.},~\cite{Khan2023,Ding2023}) are inappropriate. \textit{Finally}, this system experiences periodicity from fixed satellite orbits, and suffers uncertainties and dynamics from wireless channel conditions. How to effectively discern the periodicity and deal with the dynamics in such systems are also imperative technical challenges. As such, addressing these challenges necessitates an innovative method absent from the current literature.

\par Accordingly, we aim to propose a novel DCB online multi-objective optimization approach that is more effective than existing work. The main contributions of this paper are summarized as follows:

\begin{itemize}
 
    \item \textit{DCB-based Terminal-to-satellite Uplink Communication Systems:} We utilize DCB to enable and extend direct uplink communications between the terminals with coarse antenna and LEO satellites. This system can enhance the transmission gain of the terrestrial terminals, and thus enhance the uplink achievable rate and reduce the satellite switching frequencies of terrestrial terminals. To the best of our knowledge, such a joint optimization of satellite switching and DCB in satellite networks has not yet been investigated in the literature. 

    \item \textit{Long-term and Multi-objective Optimization Problem (MOP):} We model the aforementioned system to explore its periodicity and dynamics. Our major finding is that the total terminal-satellite uplink achievable rate, total energy consumption of terminals, and satellite switching frequency are crucial objectives that conflict with each other. Accordingly, we perform a multi-objective optimization analysis and formulate an MOP to simultaneously optimize these concerned metrics. Then, we demonstrate that this MOP is non-convex and long-term, and requires a method with enhanced portability. 
    
    \item \textit{Innovative Multi-objective Deep Reinforcement Learning (DRL)-based Solution:} Offline optimization methods are incapable of achieving the long-term optimum for this problem, while traditional reinforcement learning algorithms lack adaptability to different scenarios. To overcome this issue, we first reformulate the problem into an action space-reduced and universal multi-objective Markov decision process (MOMDP) to enhance its portability. Then, we introduce an evolutionary multi-objective DRL (EMODRL) algorithm and eliminate low-value actions to enhance its convergence performance. This algorithm is able to obtain multiple policies that represent different trade-offs among multiple objectives to accommodate diverse scenarios.

    \item \textit{Simulation and Performance Evaluation}: Simulation results demonstrate that the proposed EMODRL algorithm outmatches various baselines. Moreover, we find that DCB enables terminals that cannot reach the uplink achievable threshold to achieve efficient direct uplink transmission. In addition, it reveals that the proposed algorithm achieves multiple policies favoring different objectives and achieving near-optimal uplink achievable rates with low switching frequency.
    
\end{itemize}

\par The rest of this paper is organized as follows. Section \ref{sec:related_works} reviews the related research activities. Section \ref{sec:models_and_preliminaries} presents the models and preliminaries. Section \ref{sec:problem_formulation_and_analysis} formulates the optimization problem. Section~\ref{sec:solution} proposes the multi-objective DRL-based solution. Simulation results are presented in Section \ref{sec:simulation_results_and_analysis}. Finally, the paper is concluded in Section \ref{sec:conclusion}.

% Section
% Related Works
%
\section{Related Works}\label{sec:related_works}

\par In this work, we aim to propose a novel DCB-based terminal-to-satellite uplink communication method. This topic involves switching and handover in satellite networks and DCB optimization. Thus, we briefly introduce the related works of them as follows. 

\par \textit{Switching and Handover in Satellite Networks} In LEO satellites, the handover and switching schemes considering their mobility have been studied in previous literature. For instance, Wang \textit{et al.}~\cite{Wang2023} proposed a handover optimization strategy based on a conditional handover mechanism to enhance service continuity in LEO-based non-terrestrial networks, in which an optimal target selection algorithm was designed to the maximum reward for each conditional handover mechanism. Moreover, Song \textit{et al.}~\cite{Song2023} proposed a channel perceiving-based handover management strategy to optimize the utilization of channels and dynamically adjust the data allocation strategy in space-ground integrated information networks. Nonetheless, the aforementioned studies concentrated on the timing and strategies of satellite switching and handover, and overlooked the opportunity to augment satellite connection duration by optimizing the transmission gain of terrestrial devices.

%   Lee \textit{et al.}~\cite{Lee2023} proposed a handover protocol to address the persistent challenge of long propagation delays in LEO satellite networks, and they considered a deep reinforcement learning method to skip the measurement report in the handover procedure by leveraging its predictive capabilities. In~\cite{Zhang2021}, the authors formulated the network-flow model of the satellite switching according to the flow matrix, in which the optimal matching relationships are obtained by minimum cost and maximum flow of the network flows. In addition, the authors in~\cite{Li2020} proposed switching and handover strategies for gateway stations and ultra-dense users to maximize the overall communication quality and balance the load of satellite networks, respectively.

\par \textit{DCB Optimization:} DCB has improved the transmission performance of various distributed systems, \textit{e.g.}, Internet-of-things (IoTs)~\cite{Jayaprakasam2017}, mobile wireless sensors~\cite{Wang2021}, and automated guided vehicles~\cite{Zhang2022}. Recently, UAVs and other aerial vehicles have incorporated DCB to enhance the efficacy of air-to-ground and air-to-air communications. Leveraging their three-dimensional (3D) mobility, UAVs can dynamically navigate to locations conducive to optimal DCB implementation and adjust communication parameters to fulfill diverse objectives. As such, prior research has explored the integration of DCB in UAV networks for purposes such as secure relay~\cite{Sun2022}, confidential data transmission~\cite{Li2023}, data harvesting and dissemination systems~\cite{Li2023a}, and others. Nevertheless, the aforementioned methods are not suitable for the considered scenario since they do not consider the periodic characteristics inherent in satellite networks, and also cannot address the trade-off between satellite switching and the transmission gain facilitated by DCB.

\par Thus, different from the existing works, we consider utilizing DCB to augment both the duration and transmission gain from terrestrial terminals to LEO satellites. Based on this, we seek to devise the switching and beamforming strategies of such systems to facilitate efficient terrestrial-to-satellite uplink transmission.

% Section
% System Models and Preliminaries
%
\section{System Models and Preliminaries} \label{sec:models_and_preliminaries}

\begin{figure}
    \centering
    \includegraphics[width=1\linewidth]{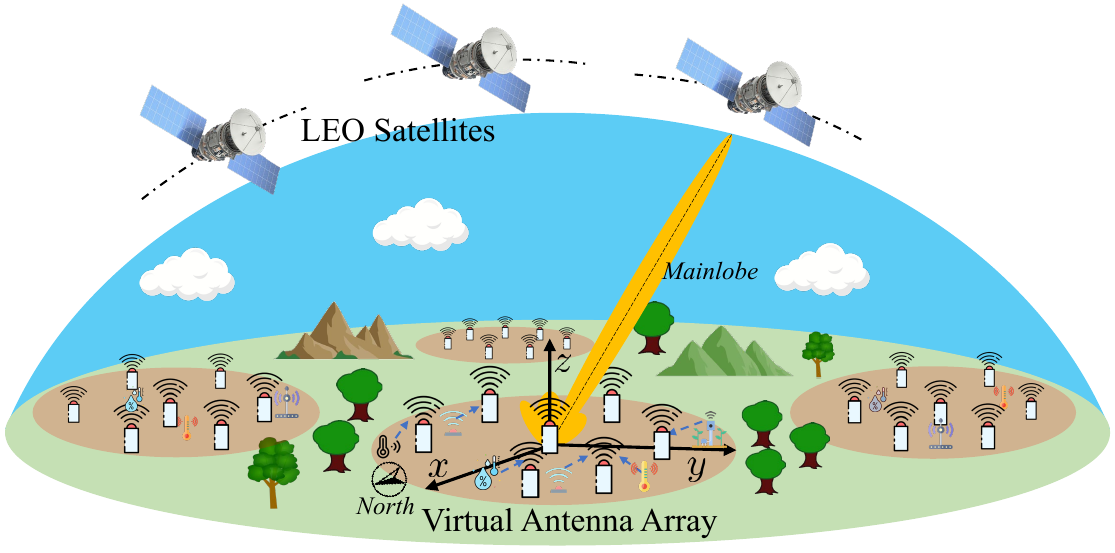}
    \caption{A terminal cluster to LEO satellites communication system. All the terminals can directly connect with LEO satellites that are with fixed earth orbits. Terminals will form a virtual antenna array and select a suitable LEO to perform uplink data transmission.}
    \label{fig:model}
\end{figure}

\subsection{Network Segments}

\par The terrestrial-satellite system under consideration is illustrated in Fig.~\ref{fig:model}, and it comprises the following elements:

\begin{itemize}

    \item A satellite network consisting of a constellation of LEO satellites $\mathcal{L}=\{\ell|1, 2, ..., N_L\}$. Each satellite may receive contents from terrestrial satellite terminals in its coverage and then transmit them to a data fusion center. These satellites are furnished with high-performance antennas with sufficient transmit power, and thus the downlink communications from satellites to terminals are efficient~\cite{RahmatSamii2015}.
    
    \item A terrestrial cluster comprising randomly distributed terminals. We consider that the geographical conditions (\textit{e.g.}, long intermediate distances, mountains, buildings, or other clustering methods~\cite{Shahraki2021}) naturally divide a large area into multiple ad hoc network clusters. Due to the link distance and channel conditions, intra-cluster communications are efficient, while the cooperation across clusters is unfeasible. These clusters may have varying numbers and distributions of terminals~\cite{Xu2017}. Thus, our primary focus is to investigate one of these clusters and propose a universal method which is applicable to such types of clusters. Without loss of generality, the cluster deploys a series of energy-sensitive and low transmission performance terminals, denoted as $\mathcal{I}=\{i|1, 2, ..., N_I\}$. Each terminal $i \in \mathcal{I}$ is able to collect data from IoT devices in coverage and needs to access the satellite network for data uploading. Due to constrained transmission resources, these terminals face challenges in establishing effective terrestrial-satellite uplinks, especially when the LEO satellite is remote.
    
    % A ground cluster with randomly distributed devices. Specifically, a series of energy-sensitive satellite terminals, denoted as $\mathcal{I}=\{i|1, 2, ..., N_I\}$, are deployed in an area, where their intra-cluster communications are feasible owing to the short link distance and favorable channel conditions. Each terminal $i \in \mathcal{I}$ is able to collect data from IoT devices in coverage and access the satellite network for data uploading. Due to constrained energy resources and limited antenna capabilities, these terminals face challenges in establishing effective links when the LEO satellite is at a considerable distance.
    
\end{itemize}

% ...\footnote{Here can add features like terminal distribution, how to communicate with each other using what fashion...}.\footnote{terminal antenna can be described in the next subsection.}\footnote{should check if I write them correctly.}

\par Terrestrial-satellite links are affected by the elevation angle of the LEO satellite. Specifically, angles that are closer to $90^{\circ}$ result in shorter terminal-to-satellite distances, increasing the probability of a line-of-sight (LoS) connection. Conversely, angles below a certain angle (\textit{e.g.,} $10^{\circ}$ in S-band scenarios or $40^{\circ}$ in Ka-band scenarios) are unable to support data uploading~\cite{3GPP38821}.

% \footnote{\color{blue} Our scenario can be like: There are some small sensors distributed in the rural area (no ground network). A series of satellite terminals(VAST, Ka-band) are deployed in this area, providing access to these small IoT sensors. The sensor data are collected by satellite terminals and are transmitted to the satellites then to the data center. So we can operate CB among satellite terminals to enhance the data transmission.}
%\par As shown in Fig.~\ref{fig:model}, we consider a heterogeneous IoT cluster to LEO satellite communication system. Specifically, some heterogeneous IoT devices are denoted as $\mathcal{I}=\{i|1, 2, ..., N_I\}$, and devices need to transmit data to the data fusion center through the LEO satellite networks. To better match the actual situation, we assume that these IoT devices are heterogeneous and equipped with an omnidirectional antenna. Then, we consider that there exists a set of LEO satellites denoted as $\mathcal{L}=\{l|1, 2, ..., N_L\}$ with a fixed earth orbit which is available to IoT devices. 

\par We assume that each terminal can access a maximum of one LEO satellite at a time. Due to their insufficient transmit power, the terminals will form a virtual antenna array to obtain a higher gain. To maximize the uplink achievable rate and duration, we assume that the virtual antenna array introduces all the terminals within a cluster. Without loss of generality, we consider a discrete-time system evolving over timeline $\mathcal{T} = \{t|1,2,..., T \}$. At each time slot, only a subset of the LEO satellites have enough spectrum resources and suitable angles to receive data from the virtual antenna array. The available LEO satellite set at $t$th time slot is denoted as $\mathcal{L}_t \subseteq \mathcal{L}$. As such, the virtual antenna array needs to select one LEO satellite to connect and we denote the index of the connected LEO at the $t$th time instant as $s_t$. Note that we assume that the mainlobe of the virtual antenna array can track the motion of the connected satellite during the time slot.

\par We also consider a Cartesian coordinate system, where the locations of the $i$th terminal and the connected LEO satellite $s_t$ at the $t$th time slot are represented as $[x^{I}_{i},y^{I}_{i},0]$ and $[x^{S}_{s_t}, y^{S}_{s_t}, z^{S}_{s_t}]$, respectively.

\par As such, the fixed communicable angles between terminals and satellites, coupled with the inherent orbital trajectories of satellites, introduce a certain periodicity to the system. Meanwhile, the limited spectral resources of satellites contribute to the uncertainty of availability, which brings dynamics to the considered system. In the following, we model the LEO satellite orbits and the communication process between the virtual antenna array and satellites to characterize the periodicity and dynamics within the system. 

%
% LEO Satellite Orbit
%
\subsection{LEO Satellite Orbit}

\par LEO satellites are a category of satellites that orbit Earth at relatively low altitudes, typically ranging from approximately 160 to 2000 kilometers. These satellites complete one orbit around Earth in a relatively short period. Mathematically, the orbit of such LEO satellites can be determined by a tuple $<\iota, \Omega, \omega, \varepsilon, \varrho, \nu>$~\cite{Montenbruck2002}, which is detailed as follows: 

\begin{itemize}
  \item \textit{Inclination Angle ($\iota$):} This angle represents the intersection between the orbital plane and the equator. In particular, an inclination angle exceeding $90^\circ$ indicates that the satellite's motion is in the opposite direction to that of Earth's rotation.

  \item \textit{Right Ascension of Ascending Node ($\Omega$):} This is the angle between the vernal equinox and the intersection of the orbital and equatorial planes.

  \item \textit{Argument of the Perigee ($\omega$):} This angle is measured between the ascending node and the perigee, which is the point where the satellite is closest to Earth, along the orbital plane.

  \item \textit{Eccentricity ($\varepsilon$):} This parameter denotes the eccentricity of the orbital ellipse. 

  \item \textit{Semi-Major Axis ($\varrho$):} This is a fundamental parameter used to describe the size and shape of an elliptical orbit. In the context of orbital mechanics, it is half of the length of the major axis, which is the longest diameter of the elliptical orbit. 

  \item \textit{True Anomaly ($\nu$):} This is the geocentric angle between the perigee direction and the satellite direction.
\end{itemize}

\par For the sake of simplicity and easy-to-access insights, we assume that the orbits of the LEO satellites are circular~\cite{Deng2021}. As such, the eccentricity ($\varepsilon$) is set to 0 and the semi-major axis ($\varrho$) is equal to the radius of the orbit $H_\ell$. Likewise, due to the circular orbit, $H_\ell = h_\ell+R_e$, in which $h_\ell$ is the altitude of satellite $\ell$ and $R_e$ denotes the radius of Earth. In this case, the angular velocity $\varpi_\ell$ of this LEO satellite is given by $\varpi_\ell= \sqrt{G M_e/H_\ell^3}$, where $G$ is the gravitational constant, and $M_e$ is the mass of Earth. Following this, the orbital period $\tau_\ell$ can be calculated as $\tau_\ell = 2 \pi / \varpi_\ell$. 
% \begin{equation} \label{eq:varpi}
% \varpi_l= \sqrt{\frac{G M_e}{H_l^3}},
% \end{equation}

\par By considering the discrete-time system, the timeline $\mathcal{T}$ can be divided into multiple time slots with length $\Delta T$. During different time slots, $\omega_{\ell}^{t} = \omega_\ell^{init} + (t \varpi_\ell\mod \tau_\ell)$ varies over time while other orbital parameters are fixed. Accordingly, let $<\iota_\ell, \Omega_\ell, \omega_\ell^t, \varepsilon_\ell, \varrho_\ell, \nu_\ell>$ be the instantaneous orbital parameters of LEO satellite $\ell$, the corresponding 3D Cartesian coordinate ($x^S_{\ell,t}, y^S_{\ell,t}, z^S_{\ell,t}$) in time slot $t$ can be given by 
\begin{equation} \label{eq:orbit}
\begin{aligned}
&x^{S}_{\ell,t} = H_\ell \left(\cos (\omega_{\ell}^{t}+\nu_\ell) \cos \Omega_\ell-\sin (\omega_{\ell}^{t}+\nu_\ell) \cos \iota_\ell \sin \Omega_\ell \right), \\
&y^{S}_{\ell,t} = H_\ell \left(\cos (\omega_{\ell}^{t}+\nu_\ell) \sin \Omega_\ell+\sin (\omega_{\ell}^{t}+\nu_\ell) \cos \iota_\ell \cos \Omega_\ell \right), \\
&z^{S}_{\ell,t} = H_\ell \left( \sin (\omega_{\ell}^{t}+\nu_\ell) \sin \iota_\ell \right), \\
\end{aligned}
\end{equation}

\noindent As can be seen, the position of a LEO satellite is regularly changed with its orbital period $\tau_\ell$ according to its orbital parameters. As such, we can learn and exploit this feature when controlling key decision variables of the system.

\subsection{Virtual Antenna Array Model}

\par In the virtual antenna array, all the terminals collaborate as one transmitter to send the same signals $s$. By simulating traditional beamforming in array antennas, their emitted electromagnetic waves will be superposed at the LEO satellite, thereby achieving additional transmission gain. To this end, we consider that the terminals perform data sharing by using the existing methods in~\cite{Feng2010, Feng2013}, which have been demonstrated to have negligible costs. Moreover, aiming at making the signals precisely superposed at the LEO satellite, the terminals within the virtual antenna arrays are synchronized in terms of the time and initial phase via the synchronization methods in~\cite{Mohanti2022,Alemdar2021}. 

\par As such, the sent signals $s$ are influenced by the characteristics of the channel between the terminals and LEO satellites. Specifically, we consider a remote rural scenario with no massive buildings that cause reflections and scattering. Moreover, due to the height of the satellite, the scattered signals cannot reach distant LEO satellites. In this case, we consider the channel model between the terminals and the satellites to be dominated by LoS. Thus, we introduce a channel model incorporating LoS path loss alongside random phases which may originate from the Doppler shift, device circuits, and other factors~\cite{Shi2023,Feng2022}. Accordingly, the channel coefficient between the terminal $i$ and satellite $\ell$ at any given time slot $t$ can be expressed as:
\begin{equation}
    h_{i, \ell}(t)=\sqrt{\beta_0 d_{i, \ell}^{-\alpha}} e^{j \psi_{i, \ell}(t)},
\end{equation}

\noindent where $\beta_0$ represents the channel power gain, $d_{i, \ell} = \sqrt{(x^{S}_{\ell,t}-x^{I}_{i})^2 + (y^{S}_{\ell,t}-y^{I}_{i})^2 + (z^{S}_{\ell,t}-z^{I}_{i})^2}$ is the propagation distance, $\alpha$ is the path loss exponent, and $\psi_{i, \ell}(t)$ denotes the channel phase shift at time slot $t$. We assume that the terminals can detect the transmitted signals from the LEO satellites and obtain the quantized version of the actual channel state information via the method in \cite{Ahmad2022}, so that quantizing the estimated channel phase shift online with the traditional channel estimation methods~\cite{Zeng2015}. 

\par Following this, as for any time slot $t$, the transmitted signal of terminal $i$ is assumed as a circularly symmetric complex Gaussian (CSCG) random variable with zero mean and unit variance, which is given by $\sqrt{P_i(t)}e^{j\phi_{i}(t)}s$, where $P_i(t)$ and $\phi_i(t) \in [-\pi, \pi]$ represent the transmit power and phase of terminal $i$ at time $t$, respectively. To ensure that the signal can reach the satellite and superimpose with other signals, this transmit power should exceed a minimal threshold and below maximum power, and this constraint is given by $P_{min} \leq P_i(t) \leq P_{max}$, $\forall i \in \mathcal{I}$, $\forall t \in \mathcal{T}$.

\par Recall that the connected satellite at time slot $t$ is denoted as $s_t$, the corresponding received signal is given by 
\begin{equation}\label{eq:SNR}
    y(t)=\sum_{\forall i \in \mathcal{I}} \sqrt{P_i(t) \beta_0 d_{i, s_t}^{-\alpha}} e^{j\left(\varphi_i(t)+\psi_{i, s_t}(t)\right)} s+v,
\end{equation}

\noindent where $v$ represents the additive white Gaussian noise at the connected satellite, modeled as a CSCG random variable with zero mean and variance $\sigma^2$. Recall that the terminals can perform online estimation of the channel phase shift, we assume that phase $\phi_i(t) = -\psi_{i, s_t}(t)$ to maximize the received signal power at the satellite~\cite{Feng2022}. As such, if the angle between them supports transmission, the signal-to-noise ratio (SNR) of the satellite is given by~\cite{Feng2022}
\begin{equation}
   \gamma_{SNR}(t) = \frac{\left(\sum_{\forall i \in \mathcal{I}} \sqrt{P_i(t) \beta_0 d_{i, s_t}^{-\alpha} }\right)^2}{\sigma^2},
\end{equation}

\par Following this, the achievable rate from the virtual antenna array to the connected satellite can be expressed as follows:
\begin{equation}\label{eq:transmission_rate}
    R(t) = B \log_2 \left(1 + \gamma_{SNR}(t)\right),
\end{equation}

\noindent where $B$ is the carrier bandwidth. As can be seen, the SNR and uplink achievable rate are primarily influenced by the instantaneous transmit powers of the terminals within the virtual antenna as well as the selection of the currently connected satellite at any time slot $t \in \mathcal{T}$.

%
% Satellite Switching Model
%
\subsection{Satellite Switching Model}

\par For any time slot $t \in \mathcal{T}$, the virtual antenna array needs to select one satellite to connect and upload the data. We assume that the virtual antenna array makes a decision at the beginning of each time slot whether to maintain the current satellite connection or select a new satellite connection. During this time slot, the virtual antenna array will always stay connected and automatically track the position of the satellite. We consider that the satellite divides its available bandwidth into distinct segments and allocates different bandwidths to individual receivers to mitigate interference. For the sake of simplicity, we assume that the satellite adopts the first-come first-served method, which means that once the allocated bandwidth is depleted, the satellite transfers to an unavailable state. This condition is clearly random to a virtual antenna array and thus modeled by a Bernoulli distribution with the probability $p$ $(0 < p < 1)$~\cite{Du2017,Lin2016}. In this case, we let $\boldsymbol{S} = \{s_t | t \in \mathcal{T}, s_t \in \mathcal{L} \}$ denote the index of the selected satellite at the timeline $\mathcal{T}$. This decision sequence variable could determine the uplink achievable rate and satellite switching frequency. 

% \par At each time slot, only a part of the LEO satellites have enough spectrum resources and suitable angles to receive the data from the virtual antenna array, and the available LEO satellite set at $t$th time slot is denoted as $\mathcal{L}_t \subseteq \mathcal{L}$.

% Subsection
% Problem Formulation and Analyses
%
\section{Problem Formulation and Analyses} \label{sec:problem_formulation_and_analysis}

\par In this section, we aim to formulate an optimization problem to improve the uplink transmission process of the virtual antenna array. We first highlight the main concern of the system, then present the decision variables and optimization objectives, and finally formulate a multi-objective optimization problem and give the corresponding analysis. 

\subsection{Problem Statement}

\par In this work, we organize energy-sensitive terminals into a virtual antenna array to enhance terminal-to-satellite uplink transmission performance and minimize the satellite switching frequency to mitigate ping-pong handover issues. As such, the considered system involves three goals, \textit{i.e.}, improving the total uplink achievable rate obtained by LEO satellites, reducing the total corresponding energy consumption, and reducing the number of satellite switches. 

\par At any time slot $t \in \mathcal{T}$, the terminal transmit powers used to communicate with the selected satellite determines the uplink achievable rate. As such, the satellite selection and the transmit powers of terminals are interdependent and coupled. Simultaneously, the transmit powers of the terminals also impact their energy consumption, while the sequential decision-making order of the satellite selection affects the satellite switching frequency. Thus, these optimization objectives have conflicting correlations. Accordingly, the coupling of variables and mutual influence of objectives require a multi-objective optimization formulation. The decision variables are introduced as follows. 

\par We define these decision variables and seek to jointly determine them: \textit{(i)} $\boldsymbol{P} = \left\{P_{i}(t) | i \in \mathcal{I}, t \in \mathcal{T} \right\}$, a matrix consisting of continuous variables denotes the transmit powers of terminals over time slots for performing DCB. \textit{(ii)} $\boldsymbol{S} = \{s_t | t \in \mathcal{T}, s_t \in \mathcal{L} \}$, a vector consisting of discrete variables represents the index of the selected satellite during the timeline. In what follows, we give the expression of the considered optimization objectives.  

\par \textit{Optimization Objective 1:} The primary objective is to improve the uplink achievable rate from the virtual antenna array to LEO satellites over the total timeline. As such, the first optimization objective is given by 
\begin{equation}
    f_1(\boldsymbol{P},\boldsymbol{S})=\sum_{t \in \mathcal{T}} R(t) \textit{d}t. 
\end{equation}

\par \textit{Optimization Objective 2:} When engaging in terminal-to-satellite communications, the transmit powers of the terminals directly determine their energy consumption. Given that the terminals are energy-sensitive and have limited supply energy, our second optimization objective is to minimize the total energy consumption of the terminals, which is designed as
\begin{equation}
    f_2(\boldsymbol{P})=\sum_{t \in \mathcal{T}} \sum_{i \in \mathcal{I}} P_i \textit{d}t. 
\end{equation}

\par \textit{Optimization Objective 3:} To maximize the uplink achievable rate and minimize the corresponding energy consumption, the virtual antenna array needs to select an appropriate satellite from the satellite list as the receiver. However, frequent satellite switching will lead to ping-pong handover issues and incur additional link costs. Hence, the third objective is to minimize the number of satellite switches (\textit{i.e.}, frequency). Let $N_t$ be the number of satellite switches at time slot $t$, and $N_t$ evolves as follows:
\begin{equation}
    N_{t+1}=\begin{cases}
N_t, & \text{if} \quad  s_t=s_{t+1} \\
N_t+1, & \text{if} \quad   s_t \neq s_{t+1}
\end{cases}.
\end{equation}

\noindent Following this, our third optimization objective is designed as
\begin{equation}
    f_3(\boldsymbol{P},\boldsymbol{S})=N_{\mathcal{T}}.
\end{equation}

\par According to the three optimization objectives above, our optimization problem can be formulated as follows:
\begin{subequations}
  \label{eq:formulation}
  \begin{align}
    (\mathrm{P1}): {\underset{\boldsymbol{P} = \{\boldsymbol{I},\boldsymbol{S}\} }{\text{min}}} \ & F=\{-f_{1}, f_{2}, f_{3} \},\\
    \text{s.t.} \quad \quad
    & P_{i}(t) \in [P_{min}, {P_{max}}], \quad \forall i \in \mathcal{I}, \forall t \in \mathcal{T}, \label{eq:const1}\\
    & s_t \in \mathcal{L}, \quad \forall t \in \mathcal{T}, \label{eq:const2}\\ 
    & R_t \geq \overline{R}, \quad \forall t \in \mathcal{T}, \label{eq:const3} 
  \end{align}
\end{subequations}

\noindent where Eqs.~\eqref{eq:const1} and~\eqref{eq:const2} show the constraints of transmit powers of the terminals and connected satellites, respectively. Moreover, Eq.~\eqref{eq:const3} ensures that the uplink obtains an achievable rate higher than the threshold. 

\subsection{Problem Analyses}

\par The problem ($\mathrm{P1}$) has the following properties. \textit{First}, the problem ($\mathrm{P1}$) is non-concave. This is due to the fact that its first objective function involves coupled variables comprising both continuous decision variables ($\boldsymbol{P}$) and integer decision variables ($\boldsymbol{S}$). \textit{Second}, the problem ($\mathrm{P1}$) contains long-term optimization objectives influenced by the periodicity of satellite orbits and the dynamic satellite availability status. \textit{Finally}, the problem ($\mathrm{P1}$) is an MOP with conflicting optimization objectives. For instance, under given channel conditions, improving the uplink achievable rate necessitates increasing the transmit powers of the terminals (\textit{i.e.,} $\boldsymbol{P}$), resulting in the more energy consumption. Likewise, if the transmit powers of the terminals are fixed, consistently selecting the satellite with the best channel condition and distance will increase the satellite switching frequency.

\par Hence, the problem ($\mathrm{P1}$) is a non-convex mixed-integer programming problem with a long-term optimization goal, incorporating dynamics and periodicity. This complexity renders it unsuitable for offline optimization methods such as convex optimization and evolutionary computing. \textit{Additionally}, the problem ($\mathrm{P1}$) is characterized as an MOP with conflicting objectives. The importance of these objectives varies in different applied scenarios and occasions. For instance, when the terminals are at low energy levels, the decision-maker seeks an energy-efficient deployment policy. Likewise, if the current data needed to be uploaded is large, the decision-maker prioritizes a policy that can maximize the uplink achievable rate. Thus, it is desirable to have a method that can achieve multiple policies for the decision-maker to select. \textit{Furthermore}, the status information of such systems (\textit{e.g.}, channel conditions) may not always known accurately. Thus, it is necessary to have an online and real-time response method for solving the problem. \textit{Finally}, while we have formulated a problem for one cluster with a fixed number of terminals, we also aim for the method could be easily adaptable to the clusters with varying terminal numbers with minimal modifications. Therefore, we require a method with enhanced portability.

\par In this case, DRL can be a promising online algorithm capable of learning periodicities and adapting to the dynamic~\cite{Zhao2023}. The aforementioned reasons motivate us to propose a DRL approach capable of addressing MOPs for solving the formulated problem.

% Section
% Multi-objective DRL-based Method
%
\section{Multi-objective DRL-based Method} \label{sec:solution}

\par In this section, we propose a multi-objective DRL-based method for solving the formulated problem. We begin by presenting the inherent challenges of applying traditional DRL to solve the problem. 

\begin{itemize}
    \item \textit{Lack of Portability:} In DRL, the set of available output actions is fixed, and once they significantly change, the DRL model needs to be re-trained. Thus, when utilizing DRL to solve our problem, a change in the number of terminals will mandate model retraining, which decreases its practicality and portability in real-world systems.
    
    \item \textit{Absence of Alternative Trade-off Policies:} hen dealing with multiple optimization objectives, DRL methods often combine multiple optimization objectives into one reward function according to their importance and roles. Then, DRL methods will derive one policy that is the most suitable for this reward function. In this case, decision-makers lack alternative trade-off policies to cater to various scenarios that prefer different optimization objectives. The obvious changes in the importance of optimization objectives require a redesign of the reward function and retraining of the DRL model, thereby diminishing its practicality.
    
    \item \textit{Challenges in Fast Learning and Convergence:} Due to the large number of satellites and their rapidly changing availability status, the traditional DRL algorithm may not swiftly acquire strategies and converge effectively.
    
\end{itemize}

\par Accordingly, our main focus is to ensure the availability of the trained DRL model under slight changes in the terminal number, and achieve multiple policies that can cover optimization objective importance varying. To this end, we will first transform our problem into an action space-reduced and more universal MOMDP.

\subsection{MOMDP Simplification and Formulation}

\par An MOMDP extends the Markov decision process (MDP) framework, which can be represented by a tuple $\langle \mathcal{S}, \mathcal{A}, \mathcal{P}, \boldsymbol{R}, \gamma, \mathcal{D} \rangle$. In the tuple, $\mathcal{S}$, $\mathcal{A}$, $\mathcal{P}$, $\gamma$, and $\mathcal{D}$ denote state space, action space, state transition probability, discount factor, and initial state distribution, respectively. Different from MDP, $\boldsymbol{R} = (r_1, \dots, r_m)$ in MOMDP is a reward vector, in which $r_m$ is the reward for the $m$th objectives. As such, some DRL methods modified for multi-objective optimization can combine the reward vector into one reward function in different forms and thereby obtain the corresponding policies that represent different trade-offs.

\par In general, the decision variables of an optimization problem (such as $\boldsymbol{P}$ and $\boldsymbol{S}$) will be the actions when this problem is represented as an MOMDP. Thus, the action space of the MOMDP should contain the transmit power of each terminal (\textit{i.e.,} $\boldsymbol{P}$). As aforementioned, this approach will decrease the portability of the method since the model needs to be re-trained when the number of terminals changes. Moreover, a large number of terminals may lead to an explosion in the possible combinations within the action space. In this case, we aim to transform the actions related to $\boldsymbol{P}$, so that mitigating the impact of terminal number changes within the virtual antenna array and reducing the action space. The main challenge of this task is to ensure the transformed actions are efficient and can still determine the trade-offs between the uplink achievable rate and energy consumption.

% \par When transferring our formulated MOP into MOMDP, the action space should contain the transmit power of each terminal (\textit{i.e.,} $\mathcal{P}$). As aforementioned, this approach will decrease the portability of the method since the model needs to be re-trained when the number of terminals changes. Moreover, a substantial number of terminals can lead to an explosion in the possible combinations within the action space. In this case, we aim to reduce the action space and enhance the expandability while balancing the trade-offs among different objectives. We first handle the actions concerning the transmit powers of terminals by weakening the impact of the number of terminals. 

\subsubsection{Action Transition}

\par To ensure the availability of the DRL model when terminal numbers vary, the key point is to fix the action dimension associated with the transmit powers of the terminals. To this end, we first derive the relationship between the importance of the objectives 1 and 2 with the optimal transmit powers of the terminals. Specifically, we only consider one-time slot optimization and let $a$ and $b$ be the weights of these two objectives. Then, we can give a new optimization problem as follows:
\begin{equation}
\label{eq:transmission-energy}
\begin{aligned}
    (\mathrm{P2}) \; {\underset{P_i}{\min}} \; &f_{RE} = a \rho_0 \sum_{i \in \mathcal{I}} P_i \Delta T - b \frac{\left(\sum_{\forall i \in \mathcal{I}} \sqrt{P_i(t) \beta_0 d_{i, s_t}^{-\alpha} }\right)^2}{\sigma^2} \\
    \text{s.t.} \; &P_{min} < P_i <P_{max}, \ i \in \mathcal{I},
\end{aligned}
\end{equation}

\noindent where the first term is to minimize the energy consumption of the virtual antenna array (\textit{i.e.,} $f_2$) while the second term is to maximize the SNR (SNR and achievable rate increase in tandem, and as such, the second term can be representative of $f_1$), and $\rho_0$ is a normalization parameter that puts the two terms in the same order of magnitude. As such, if we solve the problem $(\mathrm{P2})$ optimally, the instantaneous transmit powers of terminals that are the most suitable for the objective weights $a$ and $b$ can be obtained.

\vspace{0.5 mm}
\begin{lemma}\label{lemma:p2convex}
    In the considered scenarios and feasible set of $P_i$ ($i \in \mathcal{I}$), the problem $(\mathrm{P2})$ is convex. 
\end{lemma}
\vspace{0.5 mm}
\begin{proof}
    The second derivative of $f_{RE}$ shown in Eq.~\eqref{eq:transmission-energy} is given by 
    \begin{equation}\label{eq:derivative1}
        \frac{\partial^2f}{\partial P_i^2}=\frac{b \rho_0 \sqrt{\beta_0 d_{i, s_t}}  (\sum_{j=1,j \neq i}^{|\mathcal{I}|}\sqrt{P_j \beta_0 d_{j, s_t}})}{2\sigma^2} \frac{1}{\sqrt{P_i^3}},
    \end{equation}
    \begin{equation}\label{eq:derivative2}
        \frac{\partial^2f}{\partial P_i \partial P_j }=- \frac{b \rho_0 \beta_0}{\sigma^2} \frac{\sqrt{d_{i, s_t}} \sqrt{d_{j, s_t}}}{\sqrt {P_i P_j}}.
    \end{equation}

    \noindent Following this, the Hessian matrix of $f_{RE}$, denoted by $\boldsymbol{H}$, is given by
    \begin{equation}
    \begin{aligned}
        \boldsymbol{H} = \begin{bmatrix}
         \frac{\partial^2f}{\partial P_1^2} & \cdots  & \frac{\partial^2f}{\partial P_1 \partial P_{|\mathcal{I}|} }\\
         \vdots  & \ddots  & \vdots \\
         \frac{\partial^2f}{\partial P_{|\mathcal{I}|} \partial P_1} & \cdots  &  \frac{\partial^2f}{\partial P_{|\mathcal{I}|}^2}
        \end{bmatrix}.
    \end{aligned}
    \end{equation}

    \noindent As can be seen, the values on the diagonal of the matrix are always greater than zero. In our considered scenario, all terminals are deployed within a concentrated area. The maximum distance between terminals is significantly smaller than the satellite-terminal distance. Thus, the distances from the satellite to each terminal, \textit{i.e.,} $d_{i, s_t}$ ($i \in \mathcal{I}$), can be treated as equal. Moreover, the scenario involves the use of low-performance antenna terminals for satellite connection. For the signal to successfully propagate to the satellite, these low transmission performance terminals need to employ almost maximum transmit power. In such cases, $0 \ll P_{\min} \approx P_{\max}$, implying that the disparity among the transmit powers $P_i$ ($i \in \mathcal{I}$) is relatively small and can be neglected compared to other parameters shown in Eqs.~\eqref{eq:derivative1} and~\eqref{eq:derivative2}. Thus, the values on the diagonal are much larger than the values on the off-diagonal. In this case, the Hessian matrix $\boldsymbol{H}$ is positive semidefinite, and the problem $(\mathrm{P2})$ is convex.
\end{proof}
\vspace{0.5 mm}

\par Accordingly, $(\mathrm{P2})$ can be solved optimally or near-optimally by solvers. Consequently, the instantaneous transmit powers of terminals can be well-determined according to the objective weights $a$ and $b$. In this case, we can use the fixed-dimension weights $a$ and $b$ instead of the transmit powers of terminals as the actions of the MOMDP, which reduces the impact of terminal number varying. 

\par Following this, the computing resources of the considered DCB-based terminal-to-satellite uplink communication are often constrained, which needs a swift training process for flexible parameter tuning and rapid model deployment. To accelerate the training process, we discrete the weights associated with optimization objectives 1 and 2 by using equidistant discretization~\cite{Sutton2020}. As such, the DRL algorithms only need to consider a finite number of action options thereby facilitating the training speed.
% Given that the optimization problem shown in Eq.~\eqref{eq:formulation} is long-term and multi-objective, a single trade-off between objectives 1 and 2 at a certain moment cannot provide insufficient choices for achieving long-term trade-offs. Thus, this part constructs multiple alternative instantaneous trade-off schemes for further selection.
\par In particular, let $\mathcal{K}= \{(a_1, b_1), (a_2, b_2), ..., (a_{|\mathcal{K}|}, b_{|\mathcal{K}|})\}$ denote the alternative weight set, and then $a_k$ and $b_k$ can be established as follows~\cite{Sutton2020}:
\begin{equation}
    a_k = k/|\mathcal{K}|, \quad b_k = 1 - a_k, 
\end{equation}

\noindent Hence, the action concerning the transmit powers of terminals can be transformed to choose various alternative weight schemes within $\mathcal{K}$. This transformed action has a fixed dimension even if the terminal number changes and can represent the optimal or near-optimal transmit powers of terminals.

\subsubsection{MOMDP Formulation}

\par Benefiting from the simplification above, we can re-formulate the optimization problem shown in Eq.~\eqref{eq:formulation} as an action space-reduced and more universal MOMDP. The key elements of the MOMDP are given as follows: 

\begin{itemize}

    \item \textit{State Space:} We assume the terminals possess a precise timer and maintain satellite orbit data, thus acquiring accurate real-time satellite positions. Simultaneously, the log system of the virtual antenna array can store the index of the last-connected satellite. Note that we employ accurate real-time satellite positions to derive the transmit powers of terminals without inputting them into the DRL model. Consequently, the observable state at time slot $t$ of the virtual antenna array is given by
    \begin{equation} \label{eq:state_space}
        \boldsymbol{s}_t = \{t, s_{t-1}\}.
    \end{equation}
    
    \item \textit{Action Space:} As aforementioned, the virtual antenna array can select the trade-off schemes in $\mathcal{K}$ instead of using the DRL model to determine the transmit powers of terminals at different time slots. Except for that, the virtual antenna array should select one satellite to connect at any time slot. Thus, the actions that can be adopted at time slot $t$ by the virtual antenna array contain
    \begin{equation} \label{eq:action_space}
        \boldsymbol{a}_t = \{k_t, s_t\},
    \end{equation}
    \noindent where $k_t$ indicates the selected scheme from $\mathcal{K}$ at time slot $t$.
    
    \item \textit{Reward Function:} In DRL models, the environment furnishes immediate rewards after an action is performed, and then the agent adjusts its actions and learns the optimal policy according to the reward. Thus, it is essential to design a reasonable reward for enhancing the solving performance of such DRL models. To achieve long-term multi-objective optimization, the reward vector is 
    \begin{equation} \label{eq:reward}
        \begin{aligned}
            \boldsymbol{r}(t) = &[r_1(t), \, r_2(t), \, r_3(t)] \\
            = &[\rho_1 \hat{R}(t), \, -\rho_2\sum_{i \in \mathcal{I}} P_i \Delta T, \, -\rho_3 \kappa_t ],
        \end{aligned}
    \end{equation}
    \noindent where $\rho_1$, $\rho_2$, and $\rho_3$ are three normalization parameters intended to bring them into the same order of magnitude. Additionally, if $R(t) > \overline{R}$, then $\hat{R}(t) = R(t) $; otherwise, $\hat{R}(t) = 0$. Moreover, $\kappa_t$ is a parameter indicating whether the satellite changes (\textit{i.e.,} $\kappa_t=1$ denotes changes and vice versa). As can be seen, these three terms denote different objectives shown in Eq.~\eqref{eq:formulation}.
    
\end{itemize}

\par Based on this, we obtain an MOMDP in which the reward is a vector containing multiple objective rewards. Next, we aim to propose a novel multi-objective DRL method to obtain several long-term policies representing different trade-offs.

\subsection{EMODRL-based Solution}

% Algorithm
% EMODRL-ED3QN
%
\begin{algorithm}[tb]
    \normalem
    \small
    \caption{EMODRL-ED3QN}
    \label{algo:EMODRL-ED3QN}
    
    \KwIn{Number of learning tasks $N$, iteration number in warm-up $T_{warm}$, iteration number of each task $T_{task}$, evolution number $T_{evo}$}
    \KwOut{Pareto policy archive $\mathcal{A}$}
    
    \tcc{Warm-up stage}
    
    Initialize task population $\mathcal{P} = \emptyset$ and Pareto policy archive $\mathcal{A} = \emptyset$; 

    Generate $N$ evenly distributed weight vectors $\mathcal{W} = \{\mathbf{w}_1, \mathbf{w}_2, \dots, \mathbf{w}_N\}$; 
    
    Initialize $N$ enhanced D3QN policy $\{\pi_{\theta_1}, \pi_{\theta_2}, \dots, \pi_{\theta_N}\}$; 

    Generate learning task set $\mathbf{\Gamma} = \{ \Gamma_1, \Gamma_2, \dots, \Gamma_N \}$, where $\Gamma_n = \langle \mathbf{w}_n, \pi_{\theta_n} \rangle$; \\

    $\mathcal{P}' \leftarrow$ MMD3QN($\mathbf{\Gamma}, T_{warm}$)  \tcp*[r]{Algorithm~\ref{algo:MMD3QN}}

    Update $\mathcal{A}$ based on $\mathcal{P}'$ according to Pareto dominance; 
    
    \tcc{Evolutionary stage}

    \For {$e$ = $1$ to $T_{evo}$} 
    {

        $\mathcal{P} \leftarrow$ TPU($\mathcal{P}$, $\mathcal{P}'$) \tcp*[r]{Algorithm~\ref{algo:TPU}}
    
        $\Gamma' \leftarrow $ TS($\mathcal{W}$, $\mathcal{P}$) \tcp*[r]{Algorithm~\ref{algo:TS}}

        $\mathcal{P}' \leftarrow$ MMD3QN($\Gamma', T_{task}$)  \tcp*[r]{Algorithm~\ref{algo:MMD3QN}}

        Update $\mathcal{A}$ based on $\mathcal{P}'$ according to Pareto dominance; 

    }
    
    \textbf{Return} $\mathcal{A}$.
    
\end{algorithm}

\par The proposed EMODRL-based solution consists of multiple learning tasks, in which each task represents a specific trade-off among different optimization objectives. Following this, these tasks are collaboratively performed and learned by multiple agents. Through cooperation, the agents jointly converge towards Pareto optimal policies, thereby handling the formulated MOMDP. In what follows, we initially present the behavior and logic of an individual task and agent, and then delve into the cooperation of multiple learning tasks.

\subsubsection{Learning Task and Enhanced Dueling DQN Agent}

\par In the proposed EMODRL-based solution, the $n$th learning task can be represented as a tuple $\Gamma_n = \langle \mathbf{w}_n, \pi_{\theta_n} \rangle$, where $\mathbf{w}_n$ ($w_{m, n}>0$, $\sum_1^3 w_{m, n} = 1$) is a weight vector for optimization objectives and $\pi_{\theta_n}$ is the policy that seeks to achieve the best cumulative reward ($\sum\nolimits_{t \in \mathcal{T}} \mathbf{w}_n \boldsymbol{r}(t) $) under the current objective weights. 

\par We employ the dueling deep Q network (D3QN)~\cite{Ban2020} to learn the qualified policy ($\pi_{\theta_n}$). Specifically, D3QN is an extended version of DQN, and both of which are value-based reinforcement learning and utilize a neural network to store state and action information, \textit{i.e.}, Q-value ($Q_{\pi_{\theta_n}}(\boldsymbol{s}, \boldsymbol{a})$). Their primary objective is to discover the optimal policy $\pi^*_{\theta_n}$ and acquire the corresponding optimal state-action values $Q^*_{n}(\boldsymbol{s}, \boldsymbol{a})$, expressed as $\pi^*_{\theta_n}(\boldsymbol{s}) = \arg \max_a Q^*_{n}(\boldsymbol{s}, \boldsymbol{a})$. Different from DQN, D3QN defines the Q-value as the sum of the state value and the advantage values, \textit{i.e.,}
\begin{equation} \label{eq:dueling-DQN}
     Q_{\pi_{\theta_n}}(\boldsymbol{s}, \boldsymbol{a}) = V_{\pi_{\theta_n}}(\boldsymbol{s}) + A_{\pi_{\theta_n}}(\boldsymbol{s}, \boldsymbol{a}),
\end{equation}
\noindent where $V_{\pi_{\theta_n}}(s)$ represents the value of being in state $\boldsymbol{s}$, and $A_{\pi_{\theta_n}}(\boldsymbol{s}, \boldsymbol{a})$ represents the advantage of taking action $\boldsymbol{a}$ in state $\boldsymbol{s}$~\cite{Ban2020}. By separately estimating the state value and advantage values, the D3QN agent model can discern and prioritize actions more effectively, leading to improved learning and decision-making. Based on this, D3QN employs epsilon-greedy exploration during action selection. This strategy balances exploration and exploitation by selecting the action with the maximum Q-value with probability $1-\epsilon$ and choosing a random action with probability $\epsilon$.

\par However, the action space of the MOMDP encompasses some deterministic low-reward actions, and the epsilon-greedy cannot avoid such actions and may make meanless attempts. This inadequacy may hinder the D3QN agent from swiftly acquiring strategies and converging effectively. To overcome this issue, we seek to enhance the action selection strategy of D3QN. Specifically, the optimal policy is characterized by the exclusion of the unavailable satellites imposed by constraints on angle and spectrum resources from the action set $\boldsymbol{a}_t$. This is due to the fact that switching to an unavailable satellite will not get positive rewards in both the current step and future moments. Based on this, we propose a \textit{legitimate action select method} to mask such low-reward actions. Specifically, we define the legitimate action set at the $t$th time slot as $\boldsymbol{a}_t^{l}$, in which the actions of switching unavailable satellites have been excluded. Then, we propose an epsilon-greedy scheme as follows:  
\begin{equation}\label{eq:action-greedy}
\boldsymbol{a}_t = \begin{cases} 
\text{Random action in } \boldsymbol{a}_t^{l} & \text{with probability } \epsilon \\
\arg \, \max_{\boldsymbol{a} \in \boldsymbol{a}_t^{l}} Q_{\pi_{\theta_n}}(\boldsymbol{s}_t, \boldsymbol{a}) & \text{with probability } 1 - \epsilon
\end{cases} .
\end{equation}

\par Following this, we can use this exploration scheme to sample data and train for minimizing the loss function, thereby achieving the qualified network parameters $\theta_n$. The loss function is as follows:
\begin{equation}\label{eq:loss}
    L(\theta)=L_{\text {value }}+L_{\text {advantage }},
\end{equation}
\noindent where $L_{\text{value}} = 1/2 \left( V(\boldsymbol{s}; \theta_n) - V_{\text{target}} \right)^2$ and $L_{\text{advantage}} = 1/2 \left( A(\boldsymbol{s}, \boldsymbol{a}; \theta_n) - A_{\text{target}} \right)^2$, in which $V_{\text{target}}$ and $A_{\text{target}}$ are the target value and target advantage, respectively~\cite{Ban2020}. 

\par Next, we will present the learning tasks and introduce the interaction of these enhanced D3QN agents.

\subsubsection{EMODRL-ED3QN Framework}

\par In this part, we present an EMODRL-enhanced D3QN (EMODRL-ED3QN) to obtain a set of Pareto near-optimal policies by learning from the feedback of the environment. 

\begin{figure*}
    \centering
    \includegraphics[width=0.9\linewidth]{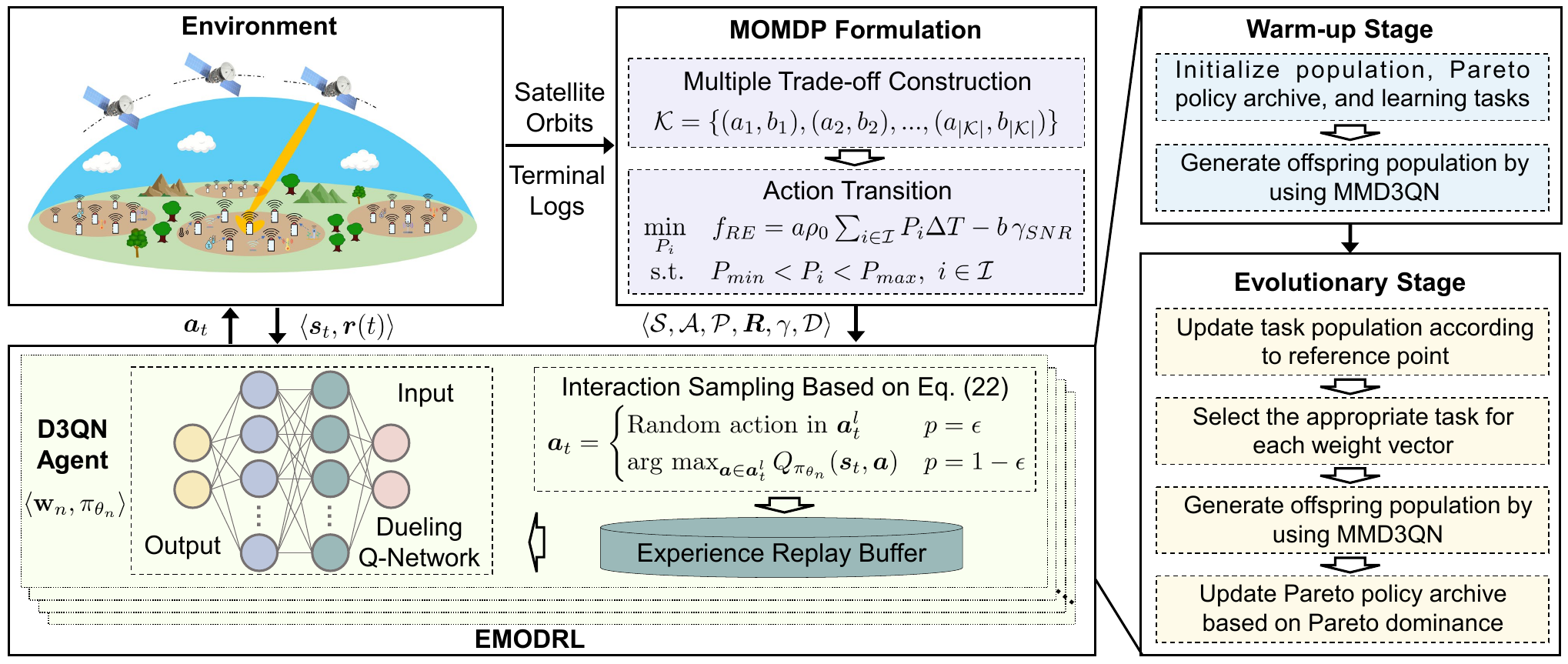}
    \caption{Framework of EMODRL-ED3QN for Multi-objective optimization in Collaborative Ground-Space Communications.}
    \label{fig:algorithm}
\end{figure*}

\par As shown in Fig.~\ref{fig:algorithm}, EMODRL-ED3QN has the same structure as the multi-objective DRL frameworks in~\cite{Song2023a,Xu2020}, which has warm-up and evolutionary two stages. In the warm-up stage, EMODRL-ED3QN generates $N$ learning tasks and generates the initial task population by using the multi-task ED3QN scheme shown in Algorithm~\ref{algo:EMODRL-ED3QN}. The evolutionary stage will update the task population, and the Pareto policy archive based on the continuously generated offspring population. These two stages are detailed as follows. 

% Algorithm
% Multi-task Enhanced Dueling DQN (MMD3QN)
%
\begin{algorithm}[tb]
    \normalem
    \small
    \caption{Multi-task Enhanced D3QN (MMD3QN)}
    \label{algo:MMD3QN}
    
    \KwIn{Task set $\mathbf{\Gamma}$, number of iterations $T_{iter}$}
    \KwOut{Offspring population $\mathcal{P}'$}

    Initialized offspring population $\mathcal{P}' = \emptyset$;

    \For {$\Gamma = \langle \mathbf{w}_n, \pi_{\theta_n} \rangle \in \mathbf{\Gamma}$}
    {
        \For{$e = 1$ to $T_{iter}$} {

            Collect data by using the proposed epsilon-greedy scheme shown in Eq.~\eqref{eq:action-greedy} \tcp*[r]{Speed up Training}

            Update network parameters by using Eqs.~\eqref{eq:dueling-DQN} and~\eqref{eq:loss};
            
        }
        
        Collect the updated new task $\Gamma_n$ in $\mathcal{P}'$;
    }
    
    \textbf{Return} $\mathcal{P}'$.
\end{algorithm}

\par $\bullet$ \textit{Warm-up Stage:} This stage stochastically generates a set of $N$ learning tasks which are defined as $\boldsymbol{\Gamma} = \{\Gamma_1, \ldots, \Gamma_N\}$. Note that these tasks share the same state space, action space, and reward vector, but have different objective weight vectors and neural network parameters. \textit{First}, the weight vectors of these tasks are assigned as $\mathcal{W} = \{ \mathbf{w}_1, \mathbf{w}_2, ..., \mathbf{w}_N\}$, in which they are evenly distributed and sampled from a unit simplex~\cite{Song2023}. \textit{Then}, we randomly initialize $N$ Q-value networks $\{Q_{\pi_{\theta_1}}, Q_{\pi_{\theta_2}}, ..., Q_{\pi_{\theta_N}}\}$. As such, $\pi_{\theta_n}$ can make decisions according to the Q-value networks and the weighted reward $\mathbf{w}_n \boldsymbol{r}(t)$. 

\par Next, we utilize the multi-task ED3QN scheme to generate the initial task population. As illustrated in Algorithm~\ref{algo:MMD3QN}, this multi-task ED3QN approach allows all learning tasks to gather data from the environment and adjust network parameters according to the main steps of the ED3QN agent. The learning tasks with the adjusted network parameters are the generated offspring task population. 

\par As such, we can obtain a set of learning tasks with well-initialized policies, and the process of the evolutionary stage can unfold as follows.

\par $\bullet$ \textit{Evolutionary Stage:} This stage explores better strategies by iteratively updating the task population. Each iteration contains three steps that are task population updating, Pareto policy updating, and offspring population generating. 

\par \uline{\textit{As for task population updating}}, we need to update the task population $\mathcal{P}$ according to the newly generated offspring population $\mathcal{P}'$ (As shown in Algorithm~\ref{algo:TPU}). In this case, it is essential to distinguish the nondominated policies and keep the population diversity. Thus, we introduce the buffer strategy~\cite{Song2023} to reasonably update $\mathcal{P}$. Specifically, multiple buffers are set to store $\mathcal{P}$, in which $B_{num}$ and $B_{size}$ are defined as their total number and capacities, respectively. As such, the objective performance space is segmented into $B_{num}$ buffers, each capable of storing up to $B_{size}$ policies. We can set a reference point $\mathbf{Z}_{ref}$~\cite{Xu2020} to prioritize these policies within the same buffer. 

\par Accordingly, for any given buffer, tasks are sorted in descending order based on their distances to $\mathbf{Z}_{ref}$. If the number of tasks exceeds $B_{size}$, only the first $B_{size}$ tasks in that buffer are retained. Following this, the learning tasks from all buffers collectively constitute a new task population.

% Algorithm
% Task Population Update (TPU)
%
\begin{algorithm}[tb]
    \small
    \normalem
    \caption{Task Population Update (TPU)}
    \label{algo:TPU}
    \KwIn{Task population $\mathcal{P}$, offspring population $\mathcal{P}'$}
    \KwOut{Updated population $\mathcal{P}$}

    Define reference point $\mathbf{Z}_{ref}$, number of buffer $B_{num}$, and size of buffer $B_{size}$;

    Initialize $B_{num}$ performance buffers $\mathcal{B}_1$, $\mathcal{B}_2$, $\dots B_{num}$;

    \For{$\Gamma = \langle \mathbf{w}_q, \pi_{\theta_q} \rangle \in \{ \mathcal{P} \cup \mathcal{P}' \}$} {

        Evaluate objective vector $\mathbf{F}(\pi_{\theta_q})$\;

        Set $\mathbf{F}_{temp} = \mathbf{F}(\pi_{\theta_q})-\mathbf{Z}_{ref}$\;

        Set index $\hat{n} = \arg \max_{n = 1, \dots, B_{num}} \{ \mathbf{w}_n \mathbf{F}_{temp} \}$\;

        Store task $\Gamma$ in $\mathcal{B}_{\hat{n}}$\;

        \If {$|\mathcal{B}_{\hat{n}}| > B_{size}$} {

            Sort all tasks in $\mathcal{B}_{\hat{n}}$ in descending order of their distances;
            
            Retain the first $B_{size}$ tasks in $\mathcal{B}_{\hat{n}}$;
        }
        
    }

    Set new task population $\mathcal{P} = \{ \mathcal{B}_1 \cup  \cdots \cup \mathcal{B}_{B_{num}} \}$\;
    \textbf{Return} $\mathcal{P}$.
\end{algorithm}

\par \uline{\textit{As for Pareto policy updating}}, a Pareto archive is utilized to retain nondominated policies discovered during the evolutionary stage. Specifically, this Pareto archive undergoes an update according to the offspring population $\mathcal{P}'$. For the ED3QN policy $\pi_{\theta}$ of each task in $\mathcal{P}'$, the policies dominated by $\pi_{\theta}$ are excluded, and $\pi_{\theta}$ is added to the Pareto archive only if no policies in the Pareto archive dominate $\pi_{\theta}$ (see step 11 of Algorithm~\ref{algo:EMODRL-ED3QN}).

\par \uline{\textit{As for offspring population generating}}, we choose the optimal task from $\mathcal{P}$ and still use the multi-task ED3QN approach to obtain the offspring task population. Specifically, we evaluate the objective function values $\mathbf{F}(\pi_{\theta_q})$ of each policy $\pi_{\theta_q}$ within $\mathcal{P}$. Then, for a given weight vector $\mathbf{w}_n \in \mathcal{W}$, we determine the best learning task in $\mathcal{P}$ based on $w_n$ and $\mathbf{F}(\pi_{\theta_q})$ ($q = 1, \ldots, |\mathcal{P}|$) (as shown in Algorithm~\ref{algo:TS}). Finally, the $N$ selected learning tasks are incorporated into $\Gamma'$. We derive $P'$ by executing multi-task ED3QN (see Algorithm~\ref{algo:MMD3QN}) with $\Gamma'$ and $T_{task}$ as its input, where $T_{task}$ represents the predefined number of task iterations.

\begin{algorithm}[tb]
    \small
    \normalem
    \caption{Task Selection (TS)}
    \label{algo:TS}
    
    \KwIn{Weight vector set $\mathcal{W}$, task population $\mathcal{P}$}
    
    \KwOut{Selected task set $\mathbf{\Gamma}'$}

    Calculate objective vector $\mathbf{F} (\pi_{\theta_n})$ of policy $\pi_{\theta_n}$ of each task $\Gamma_n \in \mathbf{\Gamma}$\;

    \For{$\mathbf{\omega}_n \in \mathcal{W}$} {
    
        Set index $\hat{q} = \arg \max_{ q = 1, \dots, |\mathcal{P}| } \{ \mathbf{w}_n \mathbf{F}(\pi_{\theta_q}) \}$\;

        Replace weight vector $\mathbf{w}_{\hat{q}}$ of $\Gamma_{\hat{q}}$ with $\mathbf{w}_i$;

        Add task $\Gamma_q$ to $\mathbf{\Gamma}'$\;
    }
    \textbf{Return} $\mathbf{\Gamma}'$.
\end{algorithm}

\par This stage terminates if the predefined number of evolution generations are completed. In this case, all non-dominated policies stored in the Pareto archive will be output as the Pareto near-optimal policies for the formulated MOMDP as well as the optimization problem. These policies represent different trade-offs between the total uplink achievable rate, total energy consumption of terminals, and total satellite switching number. As such, the decision-makers can select one policy from them according to the current requirements and concerns.

\subsubsection{Complexity Analysis}

\par We first consider the time complexity of the training EMODRL-ED3QN model. Specifically, in both the warm-up and evolutionary stages, the major complexity comes from the step of generating offspring population which involves the training of neural networks. Compared with this step, other steps (\textit{e.g.}, steps 8, 9, and 11 in Algorithm~\ref{algo:EMODRL-ED3QN}) are considered trivial and can be disregarded. 

\par As shown in Algorithm~\ref{algo:MMD3QN}, MMD3QN generates the offspring population, and its time complexity mainly depends on the training of neural networks. MMD3QN iteratively optimizes each learning task $ \pi_{\theta_n} $ in the task set for $ T_{iter} $ times (\textit{i.e.}, steps 2-8 in Algorithm~\ref{algo:MMD3QN}), where $ T_{iter} $ denotes the number of task iterations. Let $ N_{data} $ denote the number of collected data, and $ N_{epo} $ be the number of epochs for training the Q-value network. Note that the implemented Q-value network is the fully connected neural network, which consists of an input, an output, and $ C $ fully connected layers. The numbers of neurons in the input and output layers are 2 and 2, respectively. Let $ N_c$ denote the number of neurons in the $ c $th fully connected layer, with $ N_0 = 2 $ and $ N_{C+1} = 2 $. Consequently, the time complexity of MMD3QN is expressed as $ O(n \cdot (T_{iter} \cdot N_{epo} \cdot N_{data} \cdot \sum_{c=1}^{C+1} N_{c-1} \cdot N_c)) $~\cite{Song2023a}. 

\par By considering the predefined number of maximum evolution generations ($ T_{evo} $), the time complexity of training EMODRL-ED3QN is $ O(T_{evo} \cdot n \cdot (T_{iter} \cdot N_{epo} \cdot N_{data} \cdot \sum_{l=1}^{L+1} N_{l-1} \cdot N_l) $. 

\par Moreover, we analyze the time complexity of using the trained EMODRL-ED3QN. Since EMODRL-ED3QN achieves multiple alternative policies to match the current preference, using EMODRL-ED3QN does not need transfer learning or other tuning. As such, the selected policy can quickly generate a solution to the problem through simple algebraic calculations. In this case, the time complexity of using the trained EMODRL-ED3QN is $ O(T \cdot \sum_{c=1}^{C+1} N_{c-1} \cdot N_c) $, where $ T $ is the number of time slots~\cite{Song2023a}.

\section{Simulations and Analyses} \label{sec:simulation_results_and_analysis}

\par In this section, we conduct key simulations to evaluate the performance of the proposed EMODRL-ED3QN-based method for solving the formulated optimization problem.

\subsection{Simulation Setups}

\subsubsection{Scenario Settings}

\par In this work, we consider a terrestrial terminal to LEO satellite communication scenario, which includes the LEO satellite, terrestrial terminal, and communication-related parameters. \textit{First}, we set up 110 periodically operated LEO satellites, of which 80 LEO satellites at an altitude of $5 \times 10^5$ m and 30 LEO satellites at an altitude of $10^6$ m. Note that most of them are around the equatorial orbit and some of them have an inclination angle around $\pm \pi / 8$, and the satellites in the same orbit are evenly distributed in this orbit~\cite{Okati2020,Deng2021}. \textit{Second}, we consider a $100 \times 100$ terrestrial terminal area located near the equator, in which exists 10 terrestrial terminals and several sensors. Note that these devices can perform efficient information sharing and communication within the area. \textit{Finally}, the carrier frequency, minimum to maximum transmit powers of each terminal, path loss exponent, and total noisy power spectral density are set as 2.4 GHz, 1-2 W, 2, and -157 dBm/Hz, respectively. 

\par Additionally, we consider a timeline of 60 minutes. The radius of Earth $R_e$, gravitational constant $G$, and mass of Earth $M_e$ are set as $6.371 \times 10^6$ m, $6.674 \times 10^{-11} \, \text{m}^3 \, \text{kg}^{-1} \, \text{s}^{-2}$, and $5.972 \times 10^{24}$ kg, respectively. 

\subsubsection{Algorithm Settings}

\par In the proposed EMODRL-ED3QN, we set the number of the learning tasks $N$ as 10. In addition, the maximum evolution generations $T_{evo}$, the iteration number during the warm-up stage $T_{warm}$, and the iteration number for training each task $T_{task}$ are set as 300, 80, and 20, respectively. Finally, the number of performance buffers $B_{num}$ is designed to 50, and each buffer size $B_{size}$ is set to 2. For each learning task, the Q-value network has two fully connected layers with 2048 neurons and the tanh function serves as the activation function. Moreover, the learning rate is $10^{-4}$ and the discount factor is $\gamma = 0.96$. The replay buffer size and batch size are set as $10^5$ and $256$, respectively. 

\subsubsection{Baselines}

\par To demonstrate the performance of the proposed EMODRL-ED3QN, we introduce and design various comparison algorithms and strategies as follows:

\begin{itemize}

    \item \textit{Non-DCB strategy:} This strategy does not introduce DCB technology and only adopts one single terrestrial terminal to connect to the satellite directly.

    \item \textit{Achievable rate greedy policy (ARGP):} ARGP refers to the policy that any terminal $i \in \mathcal{I}$ employs the maximum transmit power $P_{max}$ and selects the satellite with the utmost uplink achievable rate at any time slot $t \in \mathcal{T}$. Note that ARGP achieves the upper bound of optimization objective 1.

    \item \textit{State-of-the-art baseline algorithms:} We design EMODRL-D3QN, EMODRL-Noisy-DQN, EMODRL-DDQN, EMODRL-PPO, EMODRL-TD3, and EMODRL-SAC as the baseline algorithms. Note that they are the variants of D3QN, Noisy-DQN~\cite{Ban2020}, double DQN (DDQN)~\cite{VanHasselt2016}, proximal policy optimization (PPO)~\cite{Schulman2017}, twin delayed deep deterministic policy gradient algorithm (TD3)~\cite{DominguezBarbero2023}, and soft actor-critic (SAC)~\cite{Haarnoja2018}, respectively. We develop them by introducing the proposed evolutionary multi-objective reinforcement learning and multi-task frameworks for dealing with the formulated MOMDP. 
\end{itemize}

\par As such, the comparison with non-DCB strategy shows the effectiveness of introducing DCB, the comparison with ARGP can assess the effect of the proposed evolutionary multi-objective DRL framework, and the comparison with other EMODRL algorithms can illustrate the optimization efficiency of EMODRL-ED3QN. In the following comparisons, we consider the average optimization objective values of these algorithms over timelines (\textit{i.e.,} $\bar{f}_1$, $\bar{f}_2$, and $\bar{f}_3$) as a performance metrics.

\subsection{Performance Evaluation}

\subsubsection{Comparisons with Non-DCB Strategy}

\par
\begin{figure}
    \centering
    \includegraphics[width=1\linewidth]{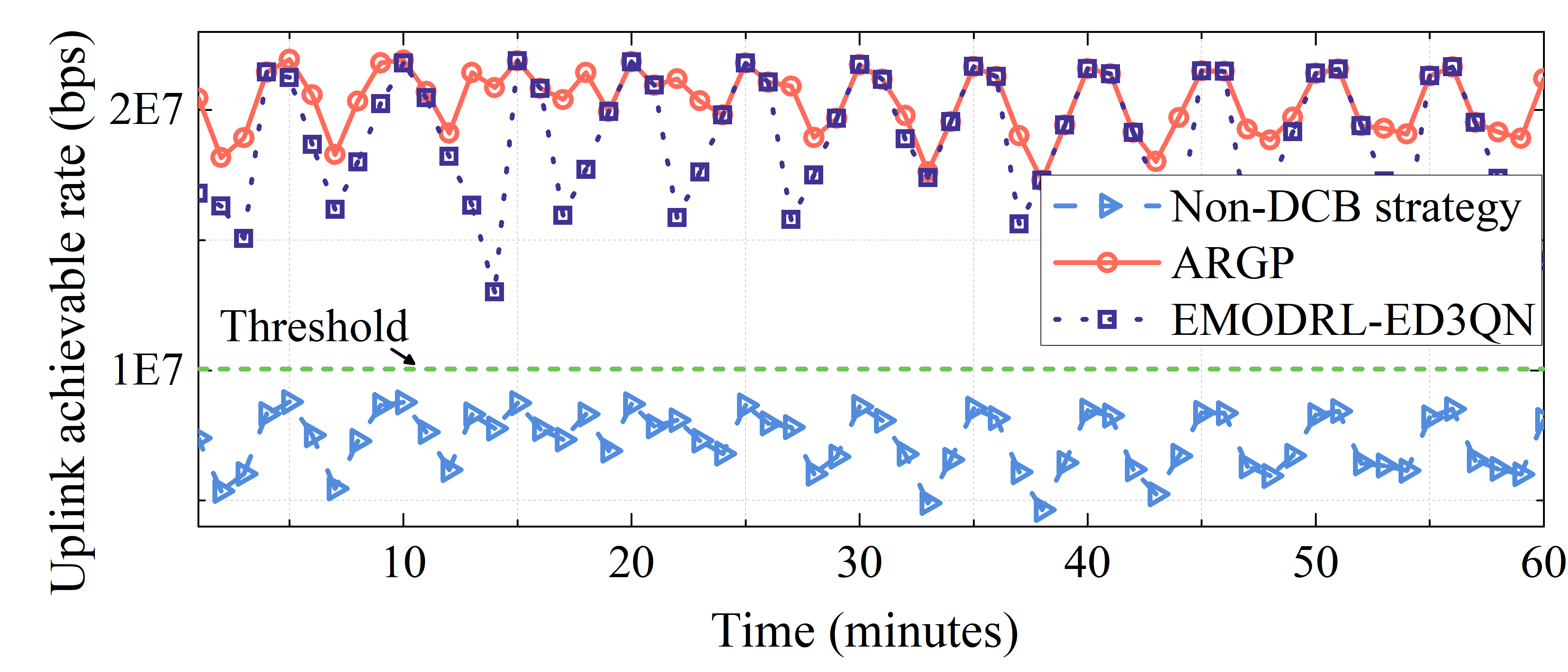}
    \caption{Uplink achievable rates obtained by an EMODRL-ED3QN policy, ARGP, and non-DCB strategy.}
    \label{fig:results}
\end{figure}

 In this part, we compare the DCB-based policies and the non-DCB strategy to illustrate the effectiveness of the considered DCB-based uplink communication approach. Specifically, uplink achievable rates obtained by a policy of EMODRL-ED3QN, ARGP, and non-DCB strategy at each time slot are shown in Fig.~\ref{fig:results}. As can be seen, ARGP and EMODRL-ED3QN consistently surpass the threshold for uplink communication. In contrast, the non-DCB strategy struggles to attain an uplink achievable rate above the threshold. Moreover, the EMODRL-ED3QN policy achieves performance closely aligned with the upper bound (\textit{i.e.}, ARGP) at every time slot. These results show that the  DCB-based uplink communication approach and EMODRL-ED3QN policy are both reasonable and suitable for the considered scenario. 

\subsubsection{Comparisons with Different Baselines}

\begin{figure}
    \centering
    \includegraphics[width=1\linewidth]{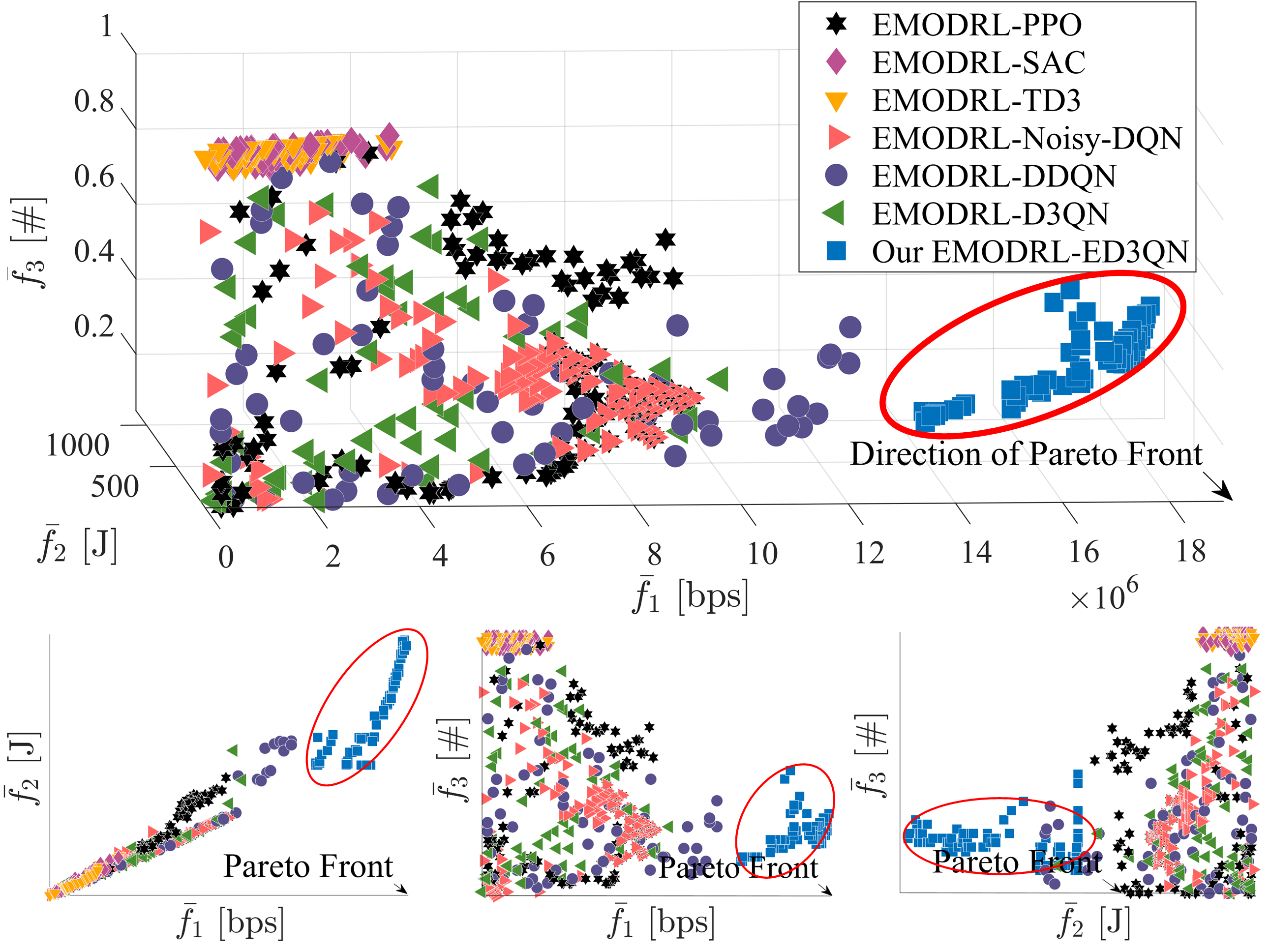}
    \caption{Pareto policy distributions obtained by different algorithms. Each point represents a Pareto policy obtained by an algorithm, and its three coordinate values represent the optimization objective values achieved by this policy. We mark the direction of the Pareto front (\textit{i.e.,} ideal Pareto policy set), and the policy closer to the Pareto front will achieve better performance.}
    \label{fig:pareto}
\end{figure}

\begin{figure*}
    \centering
    \subfloat[Satellite unavailability probability $p$]{
       \includegraphics[width=0.49\linewidth]{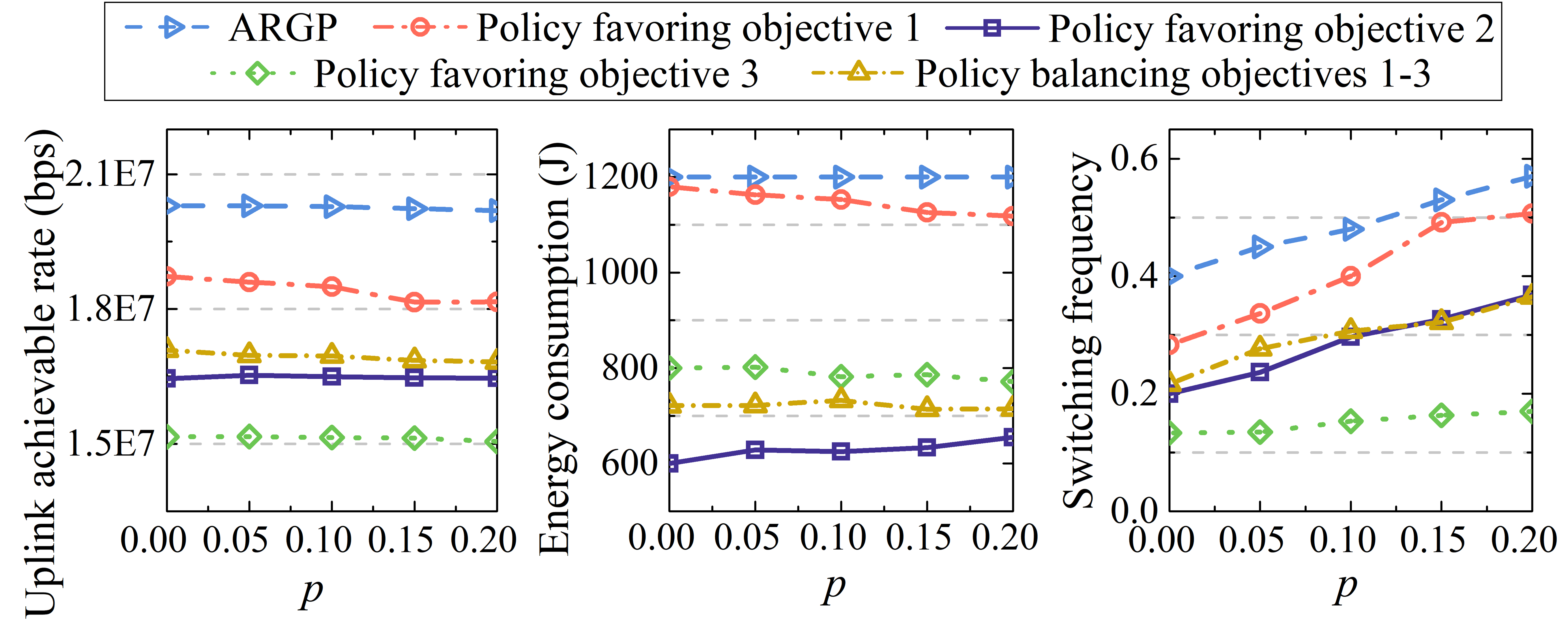}\label{fig:changep}}
    \subfloat[Terminal number]{
       \includegraphics[width=0.49\linewidth]{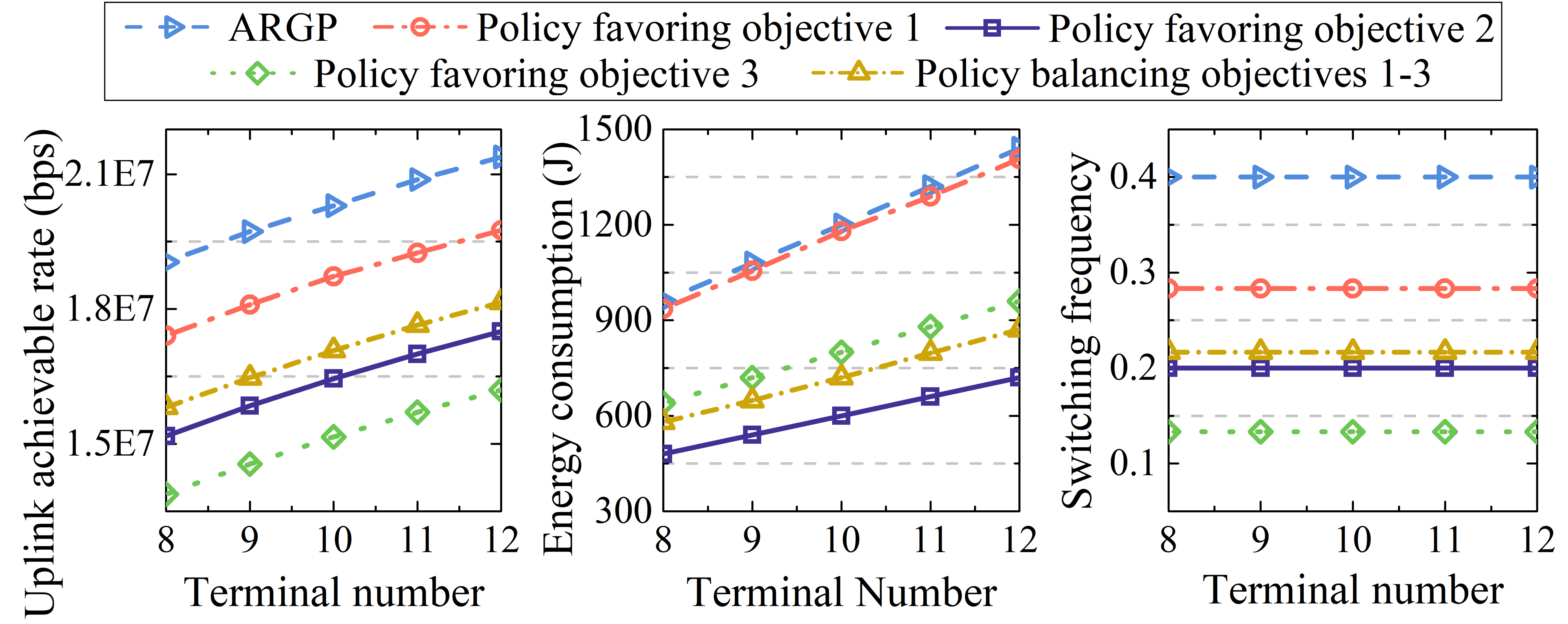}\label{fig:changenum}} 
       \\
  \caption{Impacts of scenario changes on the policies obtained by the proposed EMODRL-ED3QN.} 
  \label{fig:Impacts}
\end{figure*}

\par We first evaluate the trade-offs obtained by the proposed EMODRL-ED3QN in solving the formulated problem. As shown in Fig.~\ref{fig:pareto}, we show the trade-offs among the considered three objectives obtained by multiple EMODRL baselines. As can be seen, all these algorithms obtain a set of Pareto policies with wide coverage among the considered three objectives. Thus, the considered EMODRL framework is effective and can obtain multiple policies that weigh each other. Moreover, EMODRL-ED3QN, EMODRL-Noisy-DQN, and EMODRL-DDQN outperform other comparison algorithms. This is because the three algorithms are offline reinforcement learning methods, which may save more periodic information in their replay buffer, thereby facilitating the learning of the periodicity of the considered system. Additionally, we can see that the proposed EMODRL-ED3QN outmatches other baselines. The reason is that the proposed legitimate action select method can well-balance the exploration and exploitation of the algorithms. Moreover, the structure of the selected D3QN is also the most suitable for the designed MOMDP and legitimate action select method, and thus enables the algorithm to approach optimal performance closely.

\par Second, we select one policy from the Pareto policy set of each algorithm for further comparisons and analyses. In most cases, the uplink achievable rate from the terrestrial terminals to LEO satellites is the most concerned optimization objective. As such, we choose the policy with the best optimization objective 1 from the Pareto policy set as the final policy. In this case, the numerical results in terms of the considered optimization objectives are shown in Table~\ref{tab:result-1}. As can be seen, the proposed EMODRL-ED3QN achieves a similar objective 1 and much lower objectives 2 and 3 with the ARGP policy. This demonstrates that the proposed EMODRL-ED3QN uses lower energy consumption and satellite switching numbers to obtain a nearly optimal uplink rate. Moreover, compared with other comparison policies, EMODRL-ED3QN has a better balance among the three optimization objectives. Note that although EMODRL-PPO, EMODRL-TD3, and EMODRL-SAC achieve better optimization objectives 2 and 3, their optimization objective 1 is inadequate, making them unsuitable for terrestrial-to-satellite communication scenarios. Therefore, we can illustrate that EMODRL-ED3QN is most suitable for the considered scenario and can mitigate the ping-pong handover issue.

\begin{table}
	\centering
	\caption{Numerical results in terms of $\bar{f}_1$, $\bar{f}_2$, and $\bar{f}_3$ obtained by different baselines}
	\label{tab:result-1}
	\begin{tabular}{llll}
		\toprule
		\bf{Method} 	        & \bf{$\bar{f}_{1}$ [bps] } & \bf{$\bar{f}_{2}$ [J] }  & \bf{$\bar{f}_{3}$ [$\#$] } \\ \midrule
		ARGP              & $2.03 \times 10^{7} $          & $1200$          & $0.40$   \\
		EMODRL-PPO           & $9.33 \times 10^{6} $     & $541.84$      & $0.53$ \\   
		EMODRL-SAC          &  $3.63 \times 10^{6} $     & $174.63$      & $1.00$   \\
		EMODRL-TD3             &  $3.68 \times 10^{6} $     & $182.65$      & $0.96$   \\
		EMODRL-Noisy-DQN         & $9.74 \times 10^{6} $     & $368.24$      & $0.23$  \\
		EMODRL-DDQN             & $1.28 \times 10^{7} $     & $693.11$      & $0.36$   \\
		EMODRL-D3QN             & $1.15 \times 10^{7} $     & $613.15$      & $0.30$ \\
		\textbf{Our EMODRL-ED3QN}             & \bm{$1.87 \times 10^{7} $}     & \bm{$1179.73$}      & \bm{$0.28$} \\ \bottomrule
	\end{tabular}
\end{table}

\subsubsection{Policy Evaluations}

\par
\begin{figure}
    \centering
    \includegraphics[width=1\linewidth]{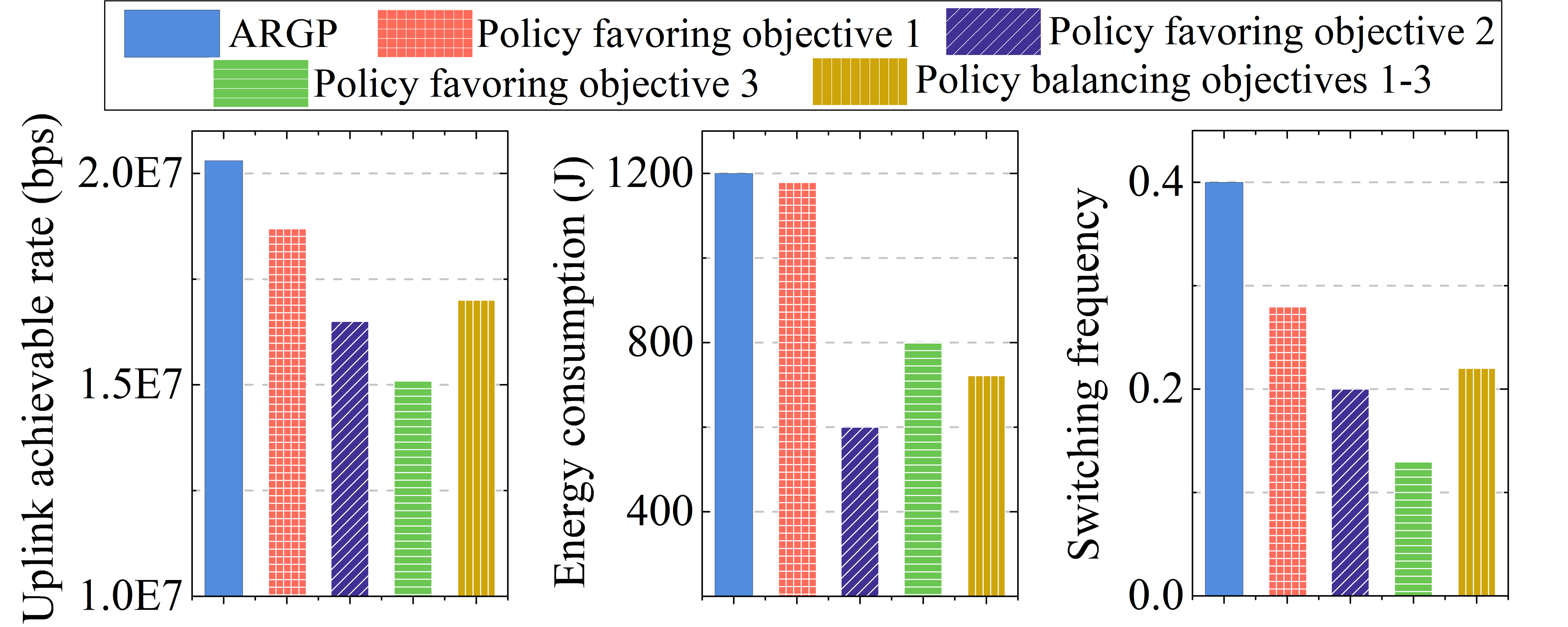}
    \caption{The optimization objective values of ARGP, policy favoring objective 1, policy favoring objective 2, policy favoring objective 3, and policy balancing objectives 1, 2, and 3.}
    \label{fig:policy}
\end{figure}

 We first select different trade-off policies from the Pareto policy archive of EMODRL-ED3QN to illustrate the diversity performance of the obtained policy set. Specifically, we select four different trade-off policies which are policy favoring objective 1, policy favoring objective 2, policy favoring objective 3, and policy balancing objectives 1, 2, and 3, and the optimization objective values of these policies are shown in Fig.~\ref{fig:policy}. It can be seen that the four policies all have obvious differences and show different objective tendencies when solving the formulated problem. In addition, these policies all achieve slightly weaker objective 1 but much better objectives 2 and 3 than ARGP. These results show that the policy set obtained by EMODRL-ED3QN has strong diversity. 

\par Then, we evaluate the impacts of scenario changes on the policies obtained by the proposed EMODRL-ED3QN. Specifically, the satellite unavailable probability may have a significant effect on these policies. Thus, we depict the changes in the three optimization objectives with satellite unavailability probability $p$ in Fig.~\ref{fig:Impacts}(a). We can observe that the policies still show different objective tendencies and no significant deterioration occurred compared with ARGP. Moreover, as aforementioned, we seek to propose a method in which one-time training can accommodate various terminal numbers. Fig.~\ref{fig:Impacts}(b) shows the performance of these trained policies changed with the terminal numbers. As can be seen, the policies still show obvious objective tendencies and achieve good performance. The reason is that the proposed legitimate action select method can enable EMODRL-ED3QN to fully explore and utilize the high-value action space and obtain more valuable trade-off policies. Thus, one-time training of the EMODRL-ED3QN can obtain multiple trade-off policies with portability.

\subsubsection{Ablation Simulations}

\par
\begin{figure}
    \centering
    \includegraphics[width=1\linewidth]{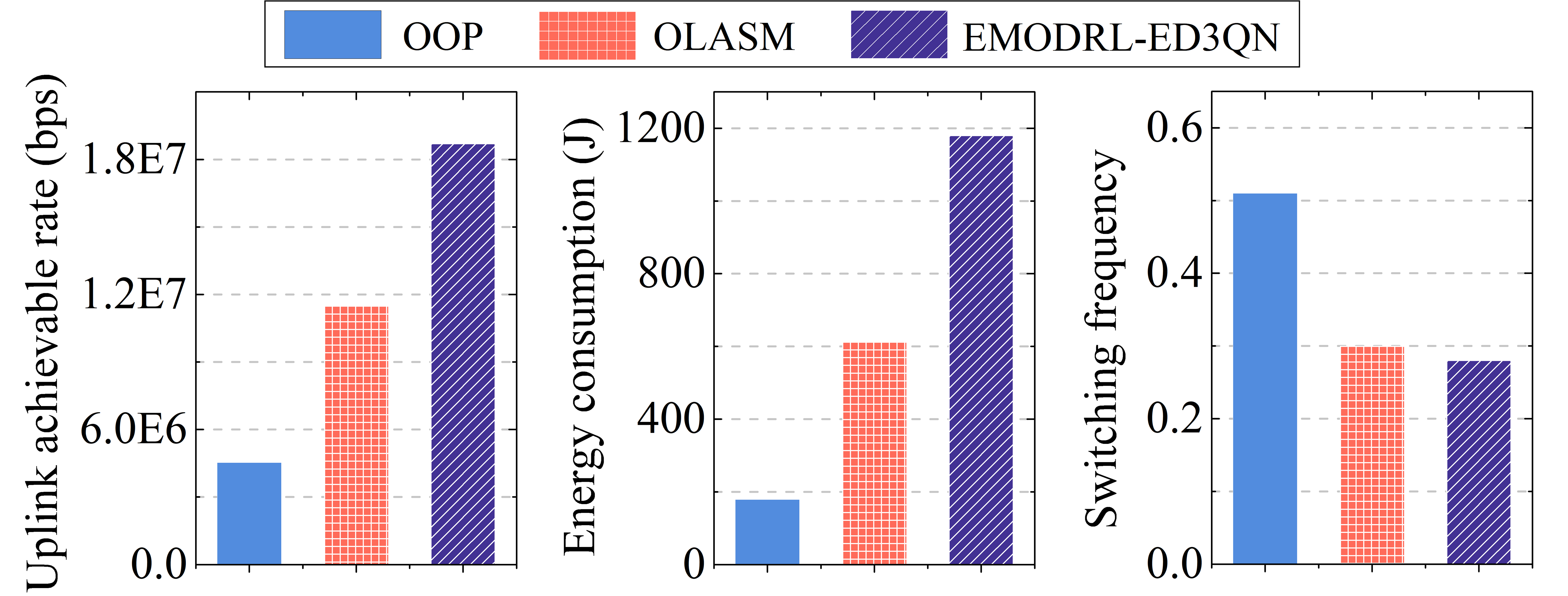}
    \caption{The optimization objective values of OOP, OLASM, and EMODRL-ED3QN.}
    \label{fig:ablation}
\end{figure}

 Ablation simulations are conducted to illustrate the effectiveness of the proposed enhanced methods. Specifically, we consider two strategies that are optimization without optimized $\boldsymbol{P}$ (OOP) and optimization without legitimate action select method (OLASM). In OOP, the transmit power of each terrestrial terminal is not optimized and randomly generated. In OLASM, the proposed legitimate action select method is not considered. Accordingly, the comparison results are shown in Fig.~\ref{fig:ablation}. As can be seen, the proposed EMODRL-ED3QN is significantly better than other ablated strategies. This shows that the proposed enhanced methods are effective and can boost the training performance of the traditional DRL algorithm in such scenarios.

% \begin{table}
% 	\centering
% 	\caption{Optimization results in terms of $\bar{f}_1$, $\bar{f}_2$, and $\bar{f}_3$ obtained by different strategies}
% 	\label{tab:result-2}
% 	\begin{tabular}{llll}
% 		\toprule
% 		\bf{Method} 	        & \bf{$\bar{f}_{1}$ [bps] } & \bf{$\bar{f}_{2}$ [J] }  & \bf{$\bar{f}_{3}$ [$\#$] } \\ \midrule
% 		OOP             & $4.53 \times 10^{6} $     & $178.66$      & $0.51$   \\
% 		OLASM             & $1.15 \times 10^{7} $     & $613.15$      & $0.30$ \\
% 		\textbf{Our EMODRL-ED3QN}            & \bm{$1.87 \times 10^{7} $}     & \bm{$1179.73$}      & \bm{$0.28$} \\ \bottomrule
% 	\end{tabular}
% \end{table}

%
%Conclusion
%
\section{Conclusion} % (fold)
\label{sec:conclusion}

\par This paper investigated a DCB-based joint switching and beamforming terminal-to-satellite uplink communication system. Specifically, we used the low transmission performance terminals as a virtual antenna array to enhance terminal-to-satellite uplink achievable rates and duration. In this system, we formulated a long-term optimization problem to improve the total uplink achievable rate, total energy consumption of terminals, and the number of satellite switches simultaneously. Following this, the problem is reformulated as an action space-reduced and more universal MOMDP to enhance its portability. Then, we proposed the EMODRL-ED3QN to obtain multiple policies that represent different trade-offs among multiple objectives to accommodate diverse scenarios. Simulation results demonstrated that EMODRL-ED3QN outmatches various baselines and obtains a wide-coverage Pareto policy set with strong usability, in which the policies achieve near-optimal uplink achievable rates with low switching frequency.

\normalem
\bibliographystyle{IEEEtran}
\bibliography{mybib}

\end{document}